# High-Temperature Anomalous Metal States in Iron-Based Interface Superconductors

Yanan Li[1,6], Haiwen Liu[2], Haoran Ji[1], Chengcheng Ji[1,9], Shichao Qi[1], Xiaotong Jiao[3,5], Wenfeng Dong[3], Yi Sun[1], Wenhao Zhang[3,8], Zihan Cui[4], Minghu Pan[5], Nitin Samarth[6], Lili Wang[3], X. C. Xie[1,9,10,11], Qi-Kun Xue[3,9,12], Yi Liu[4,7*], and Jian Wang[1,9,10‡]

[1]*International Center for Quantum Materials, School of Physics, Peking University, Beijing 100871, China*

[2]*Center for Advanced Quantum Studies, Department of Physics, Beijing Normal University, Beijing 100875, China*

[3]*State Key Laboratory of Low-Dimensional Quantum Physics, Department of Physics, Tsinghua University, Beijing 100084, China*

[4]*Department of Physics and Beijing Key Laboratory of Opto-electronic Functional Materials & Micro-nano Devices, Renmin University of China, Beijing 100872, China*

[5]*School of Physics and Information Technology, Shaanxi Normal University, Xi'an 710119, China*

[6]*Department of Physics, The Pennsylvania State University, University Park, Pennsylvania 16802, USA*

[7]*Key Laboratory of Quantum State Construction and Manipulation (Ministry of Education), Renmin University of China, Beijing 100872, China*

[8]*School of Physics and Wuhan National High Magnetic Field Center, Huazhong University of Science and Technology, Wuhan 430074, China*

[9]*Hefei National Laboratory, Hefei 230088, China*

[10]*Collaborative Innovation Center of Quantum Matter, Beijing 100871, China*

[11]*Institute for Nanoelectronic Devices and Quantum Computing, Fudan University, Shanghai 200433, China*

[12]*Southern University of Science and Technology, Shenzhen 518055, China*

‡Corresponding author. jianwangphysics@pku.edu.cn

*Corresponding author. yiliu@ruc.edu.cn

**The nature of the anomalous metal state has been a major puzzle in condensed matter physics for more than three decades. Here, we report systematic investigation and**

**modulation of the anomalous metal states in high-temperature interface superconductor FeSe films on $SrTiO_3$ substrate. Remarkably, under zero magnetic field, the anomalous metal state persists up to 20 K in pristine FeSe films, an exceptionally high temperature standing out from previous observations. In stark contrast, for the FeSe films with nano-hole arrays, the characteristic temperature of the anomalous metal state is considerably reduced. We demonstrate that the observed anomalous metal states originate from the quantum tunneling of vortices adjusted by the Ohmic dissipation. Our work offers a perspective for understanding the origin and modulation of the anomalous metal states in two-dimensional bosonic systems.**

In general, the zero-resistance superconducting state with phase-coherent Cooper pairs and the insulating state with localized Cooper pairs are believed to be the two ground states of two-dimensional (2D) bosonic systems[1,2]. However, when approaching zero temperature, a finite resistance saturation has been experimentally detected and regarded as the signature of anomalous metal state in various 2D superconducting systems, including amorphous and granular films[3-8], crystalline films and nanodevices[9-15], superconducting arrays[16-19], and interfacial superconducting systems[20,21]. Different from the conventional metals which consist of fermionic quasiparticles, the transport behavior of the anomalous metal state is dominated by the bosonic Cooper pairs, revealed by the vanishing Hall coefficient[6,19] and quantum oscillations with a period of one superconducting flux quantum $\frac{h}{2e}$ ($e$ is the electron charge and $h$ is the Planck's constant)[19]. The existence of the anomalous metal state indicates a new quantum ground state of Cooper pairs, which challenges the prevailing consensus[1,22]. Despite the ubiquitous experimental observations of anomalous metal state, the microscopic origin of this intriguing metallic ground state showing finite resistance is still poorly understood[1,22]. In particular, no general knowledge has been obtained on the evolution of this exotic anomalous metal state with increasing temperature.

In the transport characteristics of 2D bosonic systems, the vortex dynamics plays a crucial role[23]. The vortex motion can break the coherence of Cooper pairs and give rise to finite resistance. In the comparably high temperature regime, the classical motion of vortices dominates. The well-known examples are the free vortex motion above the Berezinskii-Kosterlitz-Thouless (BKT) transition temperature[24] and the thermally activated flux flow [25]. At zero field, the vortex and antivortex are unpaired above the BKT temperature, and then the free vortex motion leads to finite resistance[24]. In the thermally activated flux flow, the vortex motion over the pinning potential gives rise to the activated behavior of the resistance[25]. In the low temperature regime, the classical motion of vortices is gradually replaced by the quantum tunneling of vortices (also called quantum creep). Previous works indicate that the quantum tunneling of vortices could result in finite resistance at low temperatures under external magnetic field, which is a potential phenomenological explanation of the anomalous metal state[12,26,27]. However, this theoretical explanation is not applicable to the anomalous metal states observed under zero magnetic field[11,17-20,28]. Compared with conventional 2D superconducting systems, the ultrathin crystalline FeSe films grown on $SrTiO_3$ (STO) substrate possess high-temperature interface superconductivity with onset critical temperature above 40 K. The superconductivity emerges at the interface and is localized in the first unit-cell FeSe on STO as shown by scanning tunneling microscopy studies[29,30]. The 2D nature of superconductivity has been justified by the typical BKT transition and the strongly anisotropic critical fields[31]. The high-temperature interface superconductivity makes FeSe/STO a promising platform for studying the origin and evolution of the anomalous metal state.

In this paper, we report systematic transport measurements on crystalline FeSe films down to one unit-cell thickness grown on STO (001) substrates by molecular beam epitaxy. The scanning tunneling microscope image shows the tetragonal lattice structure of a typical FeSe/STO sample (Fig. S1). For *ex situ* transport measurements, we grew a series of macroscopic FeSe films on pretreated insulating STO substrates with FeTe protection layers (see Methods in the Supplemental Material[32]). Under zero magnetic field, the sheet resistance ($R_{\mathrm{s}}$) versus temperature ($T$) curves of FeSe films reveal a superconductor to weakly localized metal transition with increasing normal state resistance (Fig. S2). A typical superconducting FeSe/STO sample ($S1$) shows zero resistance within the measurement resolution at $T_{\mathrm{c}}^{\mathrm{zero}} = 17.4$ K and the onset superconducting critical temperature $T_{\mathrm{c}}^{\mathrm{onset}} = 46.6$ K (Fig. S2). Here, $T_{\mathrm{c}}^{\mathrm{onset}}$ is defined as the temperature where the sheet resistance deviates from the linear extrapolation of the normal state, and the normal state sheet resistance is defined as $R_{\mathrm{N}} = R_{\mathrm{s}}(T_{\mathrm{c}}^{\mathrm{onset}})$. Figure 1(a) presents the Arrhenius plots (lg$R_{\mathrm{s}}$ versus $1/T$) of sample $S2$ under different perpendicular magnetic fields ( $B_{\perp}$ ). Strikingly, the anomalous metal state, characterized by the resistance saturation, persists up to $T^{\mathrm{AM}}$ = 19.7 K (56.1% of $T_{\mathrm{c}}^{\mathrm{onset}}$) at zero field, much higher than previous reports[22,27]. An external magnetic field up to 15 T broadens the superconducting transition and reduces $T^{\mathrm{AM}}$ from 19.7 to 7.0 K [Fig. 1(a)]. The Hall ($R_{\mathrm{yx}}$) and longitudinal resistance ($R_{\mathrm{s}}$) were simultaneously measured for sample $S3$ showing the anomalous metal state [Fig. 1(b) and Fig. S6]. Above the superconducting transition, the small negative Hall coefficient indicates the heavy electron doping in the FeSe/STO, consistent with previous reports[69]. Below $T_{\mathrm{c}}^{\mathrm{onset}}$, the Hall coefficient ($R_{\mathrm{yx}}/B$, $B$ is the magnetic field) drops with decreasing temperature and reaches zero within the measurement resolution below 17 K while $R_{\mathrm{s}}$ starts to saturate below $T^{\mathrm{AM}}$ of 15.6 K, indicating that Cooper pairs (bosons) dominate the transport behavior of the anomalous metal states[6,19].

In previous studies of conventional superconductors, the anomalous metal state normally exists at hundreds of millikelvin[5,6,9,14-18,20] [see Fig. 1(c)]. In high-temperature cuprate superconductors [e.g., $La_2CuO_{4+\delta}$ and nanopatterned $YBa_2Cu_3O_{7-x}$ (YBCO) films, the open orange circles in Fig. 1(c)], although the anomalous metal state is reported up to ~10 K, the ratio of $T^{\mathrm{AM}}/T_{\mathrm{c}}^{\mathrm{onset}}$ is comparably small (0.125 for YBCO and 0.250 for $La_2CuO_{4+\delta}$). Therefore, the extremely high $T^{\mathrm{AM}}$ up to 19.7 K accompanied by the large ratio of $T^{\mathrm{AM}}/T_{\mathrm{c}}^{\mathrm{onset}}$ (0.561) in ultrathin pristine FeSe films [the open orange diamond in Fig. 1(c)] is very striking in this context. The high $T^{\mathrm{AM}}$ not only excludes the possible influence of external high frequency noise, but also makes it easier to investigate the evolution of anomalous metal state in a wide temperature regime. In addition, the large ratio of $T^{\mathrm{AM}}/T_{\mathrm{c}}^{\mathrm{onset}}$ can also be found in $LaAlO_3/KTaO_3$ (LAO/KTO) and LAO/STO interface superconductors [the blue diamond dots in Fig. 1(c)], indicating that the interface effect may enhance the anomalous metal states.

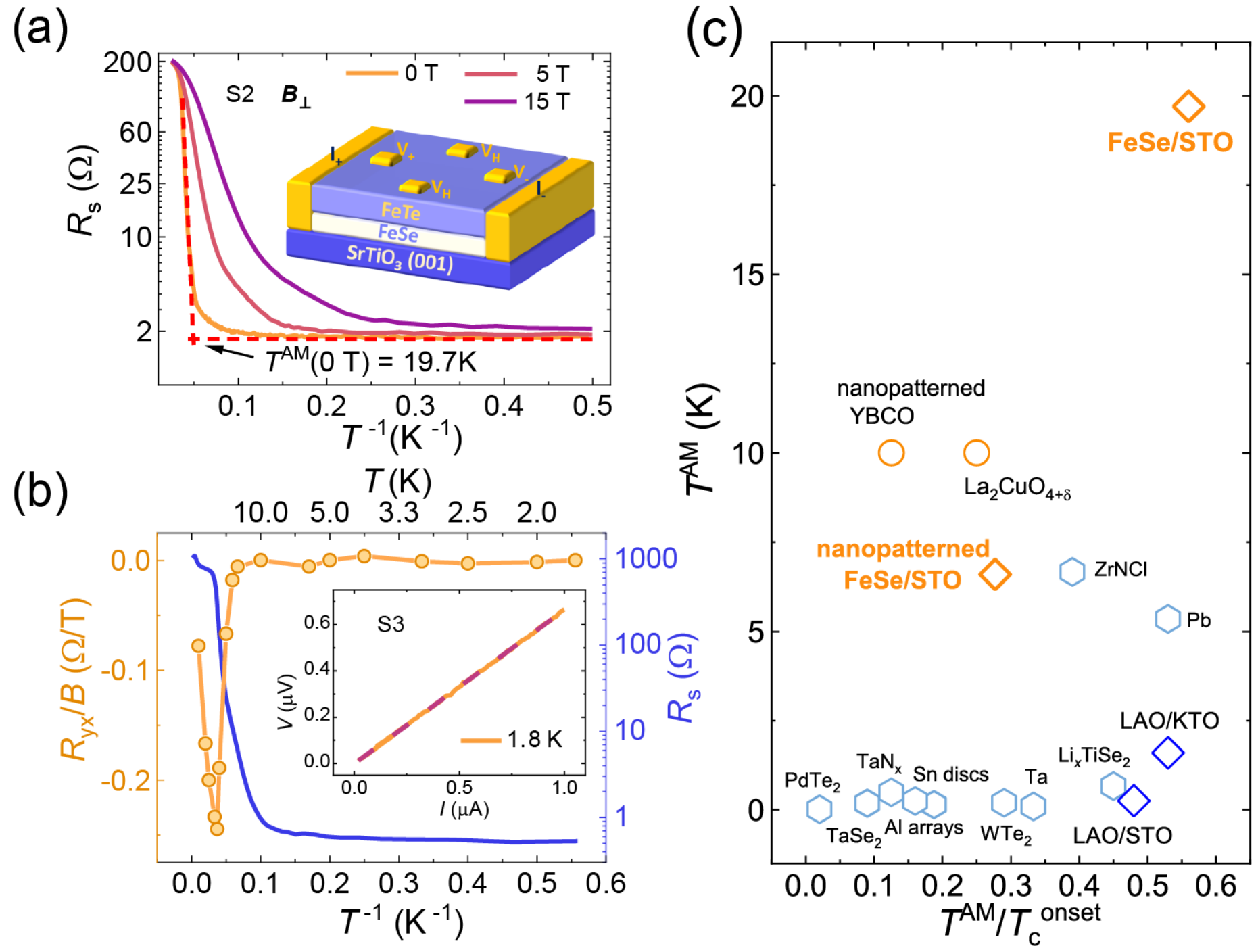

FIG. 1. High-temperature anomalous metal states in FeSe/STO. (a) Arrhenius plots of $R_s(T)$ curves under perpendicular magnetic fields of 0, 5, 15 T, showing an anomalous metal state with $T^{AM}$ up to 19.7 K at zero magnetic field (sample *S*2). $T^{AM}$ is the temperature where the anomalous metal state appears, defined as the crossing point of the extrapolation of resistance drop and saturation. Inset: the schematic for six-probe transport measurements on FeSe/STO. (b) The $R_s(T)$ curve (the blue curve) and the Hall coefficient $R_{yx}/B$ (the yellow dots) of sample *S*3. The Hall coefficient goes to zero within the measurement resolution below 17 K, comparable to the $T^{AM}$ of 15.6 K. Inset: the *I*-*V* curve at 1.8 K showing a linear behavior below 1 μA. The excitation current for $R_s(T)$ measurements is 100 nA within the Ohmic regime. (c) Overview of $T^{AM}$ and the ratio $T^{AM}/T_c^{onset}$ of various 2D superconducting systems[3,5,6,9,11,12,14,15,17-21]. The unpatterned and nanopatterned FeSe/STO in this paper are indicated by yellow diamond symbols. The highest values of $T^{AM}$ and $T^{AM}/T_c^{onset}$ from each 2D superconducting system are shown in this panel.

To further explore the nature of the anomalous metal states in FeSe/STO systems, we tune the transport properties of FeSe/STO by fabricating nano-hole arrays. Specifically, the superconducting FeSe/STO samples are etched through a contact mask via reactive ion etching (RIE) (see Methods in the Supplemental Material for details[32]). The patterns, which have ~70 nm diameter holes arranged in a triangular array with center-to-center spacing of ~102 nm, are transferred onto the FeSe films (see Fig. S8 for a scanning electronic microscope image of the patterns). The intermediate triangular superconducting areas [marked as island in Fig. 2(a)] between the nano holes connect to each other through the links, forming a Josephson junction array (JJA) of FeSe/STO [Fig. 2(a)]. A longer etching time makes the links between superconducting areas more resistive, which increases $R_{\mathrm{N}}$ and the disorder strength [Fig. S4(a)]. In Fig. 2(c), the 210 s etched film shows the anomalous metal state with a significantly suppressed $T^{\mathrm{AM}}$ of 0.4 K, nearly 2 orders of magnitude smaller than that of unpatterned FeSe/STO sample *S*2. The anomalous metal states with relatively low $T^{\mathrm{AM}}$ are also confirmed in another two nanopatterned samples [Fig. S7(c) and S7(d)]. The highest $T^{\mathrm{AM}}$ in nanopatterned FeSe/STO samples can reach 6.6 K, still lower than that in unpatterned FeSe/STO samples. Additionally, the saturated resistance increases with increasing perpendicular field [Fig. 2(c)], consistent with the giant magnetoresistance at low temperatures [Fig. S10(a)]. Furthermore, within ±0.8 T the magnetoresistance oscillates with a monotonically rising background, as shown in Fig. 2(d). The oscillation period is 0.218 ± 0.004 T, consistent with one superconducting flux quantum $\phi_0 = \frac{h}{2e}$ threading an area of one unit cell of the nanopattern (around 9010 $\mathrm{nm}^2$) in the JJA ($e$ is the electron charge). The $h/2e$ quantum oscillations persist up to 4 K [Fig. 2(d)], which demonstrates the bosonic nature of the observed anomalous metal state. Moreover, for both pristine and nanopatterned FeSe films, a remarkable linear-in-temperature ($T$-linear) resistance in a wide temperature regime below $T_{\mathrm{c}}^{\mathrm{onset}}$ is detected (Fig. 3). As shown in Fig. 3(b), with increasing etching time from 0 s to 210 s, the slope and temperature regime of the $T$-linear resistance increases significantly in the nanopatterned FeSe films. The $T$-linear resistance emerges below $T_{\mathrm{c}}^{\mathrm{onset}}$ and shows a very large slope (Fig. 3, Figs. S3 and S4) compared to the fermionic case, indicating the dominant role of Cooper pairs in this new bosonic quantum state (i.e., bosonic strange metal state, see the Supplemental Material Sec. III, for details[32]). Moreover, the $h/2e$ quantum oscillations [Fig. 2(d)] and suppressed Hall coefficient [Fig. S6(d)] are also detected in the temperature regime of $T$-linear resistance, further supporting its bosonic nature.

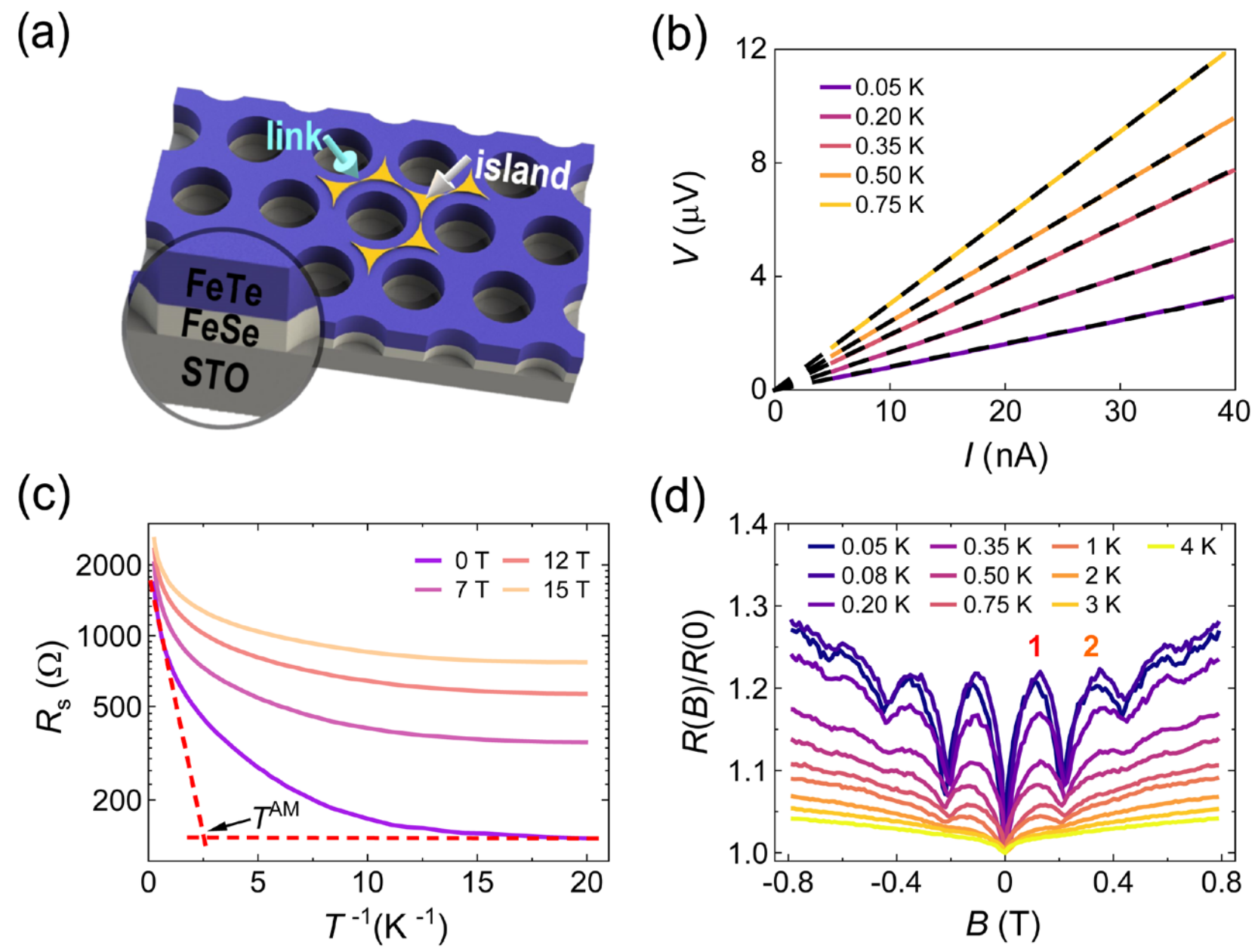

FIG. 2. The anomalous metal state in FeSe/STO with triangular arrays of nano holes (sample *S*9-210 s etch). (a) Schematic for nanopatterned FeSe film on STO. Superconducting areas (marked as island) are connected by the non-superconducting links. (b) The *I*-*V* curves showing Ohmic behavior within 40 nA down to 50 mK for 210 s etched film. (c) Arrhenius plots of $R_s(T)$ curves of 210 s etched sample under different magnetic fields, measured at 15 nA. (d) Magnetoresistance from -0.8 T to 0.8 T, showing quantum oscillations. Number 1 and 2 mark the first and second peaks of the quantum oscillations, respectively. The oscillation period is 0.218 ± 0.004 T, corresponding to one superconducting flux quantum for a unit cell pattern of $9.01\times10^3$ $nm^2$.

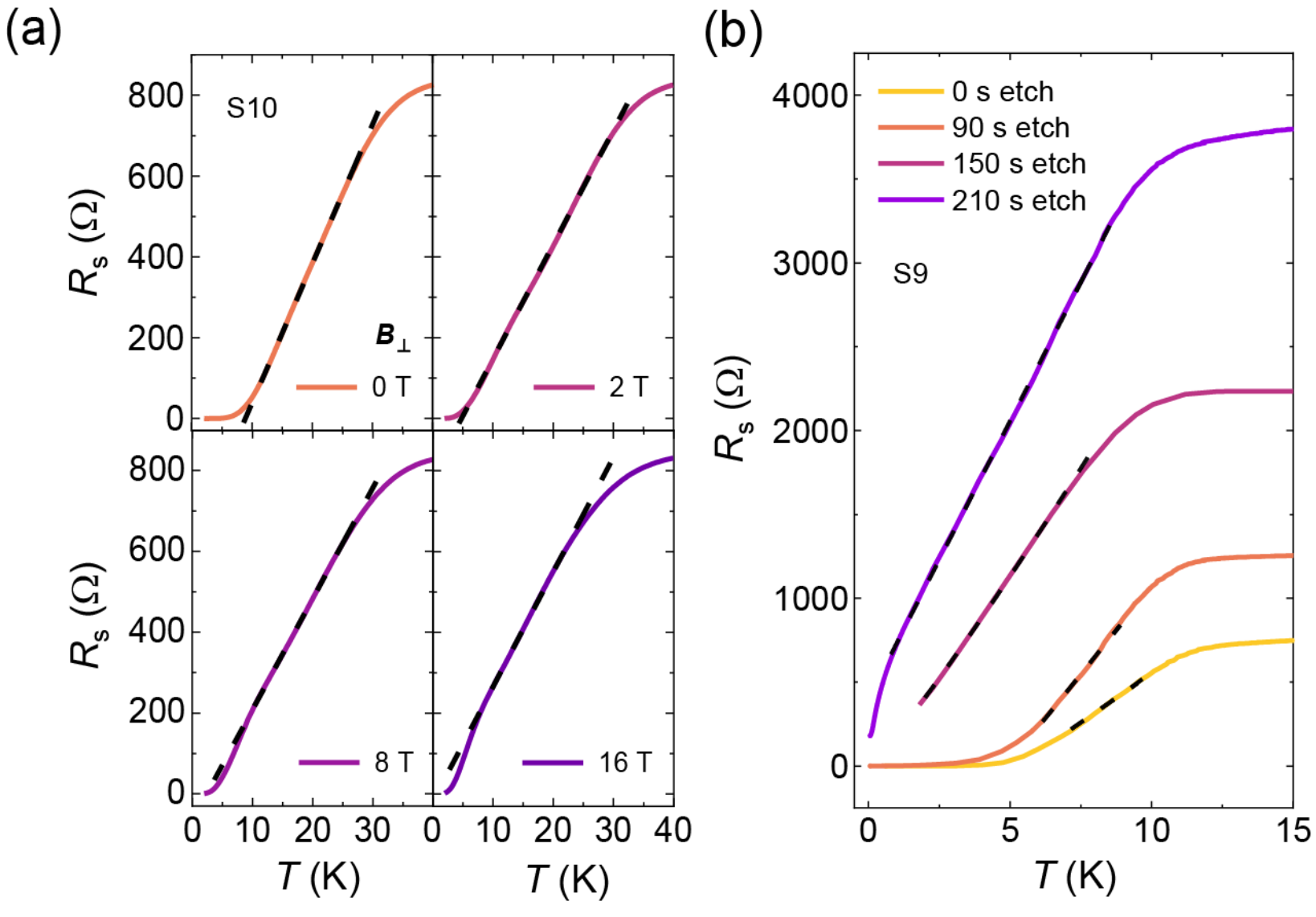

FIG. 3. The *T*-linear resistance of pristine and nanopatterned FeSe/STO. (a) The $R_s(T)$ curves of the pristine FeSe/STO (*S*10) under different perpendicular magnetic fields, showing *T*-linear resistance below $T_c^{\mathrm{onset}}$. (b) The $R_s(T)$ curves of the nanopatterned FeSe/STO (*S*9) with different etching times. A longer etching time corresponds to a larger slope and a wider temperature regime of *T*-linear resistance.

The vortex dynamics in 2D superconductors can be analyzed based on a JJA model by considering the Josephson coupling energy ($E_\mathrm{J}$) and charging energy ($E_\mathrm{c}$)[70]. JJA has been widely used as a representative system to investigate the transport features in 2D superconductors, including the demonstration of the BKT transition[71]. For the superconducting systems with finite $E_\mathrm{c}$, the quantum fluctuations can induce phase dynamics under zero magnetic field, which gives rise to the duality of Cooper pairs and vortices near zero temperature[70]. In 2D superconducting systems with weakly coupled Josephson junctions (commonly $E_\mathrm{J}$ is still larger than $E_\mathrm{c}$), the vortex dynamics plays a dominant role in the electric resistance of the system. In these systems, the bosonic modes are affected by the Ohmic dissipation generated by the coupling with fermionic modes, which represents a boson-dominated case rather than pure bosonic physics[72]. The Ohmic dissipation manifests as a friction on the vortex motion, and largely influences the vortex dynamics in the low temperature regime[73-76]. For example, in the single shunted Josephson junction model, the Ohmic dissipation strength is proportional to the inverse of the shunted resistance of the junction. A large Ohmic dissipation can pin down the time-dependent evolution of phase difference (similar concept as the vortices in JJA) and give rise to a superconducting feature, while a small Ohmic dissipation cannot pin down such motion and leads to finite resistance close to the quantum resistance of $\frac{h}{4e^2}$[74,77].

Thus, we analyze the behavior of anomalous metal states considering the quantum tunneling of vortices in JJA with the Ohmic dissipation, and the 2D dissipative quantum *XY* model provides a microscopic basis for the analysis of vortex dynamics[75,78,79]. In the moderate temperature regime, the tunneling rate of vortices is influenced by the thermal effect, and the thermal-smeared quantum tunneling rate [Fig. 4(a)] is predominantly determined by the dimensionless dissipation strength $\bar{\gamma}$ and follows a power-law temperature dependence[73,74,80]. Therefore, the sheet resistance of the superconducting system, proportional to the tunneling rate of vortices, satisfies $R_\mathrm{s} \propto T^{2\bar{\gamma}-1}$ in this regime, where $\bar{\gamma} = \frac{h}{4e^2\overline{R_L}}$ and $\overline{R_L}$ is the average link resistance of the weakly coupled Josephson junction. For our samples, the possible influence of nonsuperconducting upper FeSe layers can be included in $\overline{R_L}$ (see the Supplemental Material, Sec. V, for details[32]). In the comparably low temperature regime, the sheet resistance is contributed from the quantum tunneling of vortices [Fig. 4(a)], and follows the relation $R_s = R_0 Exp[T/T_0]$, where the saturated resistance $R_0$ is determined by the link resistance $R_L$ and $T_0$ is an intrinsic energy scale. The derivations of the tunneling rate of vortices and sheet resistance in the moderate and low temperature regime are shown in the Supplemental Material[32]. In Fig. 4(b), the experimental $R_s(T)$ curve (the orange line) of sample $S2$ showing a high-temperature anomalous metal state can be well fitted by our theoretical model (purple lines) in both moderate and low temperature regime. Moreover, the $R_s(T)$ behaviors of the anomalous metal states in eight FeSe/STO samples can be quantitatively explained by our theoretical model (Figs. S11 and S12), demonstrating that the anomalous metal state under zero magnetic field originates from the quantum tunneling of vortices influenced by the Ohmic dissipation.

The evolution of anomalous metal states with increasing normal state resistance $R_N$ during the quantum phase transition can be understood within our theoretical analysis incorporating the varying dissipation strength $\bar{\gamma}$. As shown in Fig. 4(c), our experimental study reveals that the $T^{AM}/T_c^{onset}$ decreases with increasing $R_N$ in both pristine and nanopatterned FeSe/STO samples, suggesting that $R_N$ is an experimentally accessible indicator to characterize the anomalous metal states. In the theoretical simulation, we assume the average link resistance $\overline{R_L}$ is roughly proportional to $R_N$. In Fig. 4(d), we have plotted the theoretical analysis result for $\lg(R/\widetilde{R_c})$ versus $T_c^{onset}/T$ curves ($\widetilde{R_c} = \frac{h}{4e^2} \cdot \frac{E_c}{4k_B T_c^{onset}} \cdot \frac{\sqrt{\pi}}{2}$) with different colors representing different dissipation strength $\bar{\gamma}$ ($\bar{\gamma} = \frac{h}{4e^2 \overline{R_L}}$) to reveal the contribution of quantum tunneling of vortices with Ohmic dissipation. For samples with large dissipation $\bar{\gamma}$, the sheet resistance decreases rapidly below $T_c^{onset}$, and the system quickly enters into the quantum tunneling regime with a residual resistance, suggesting a high-temperature anomalous metal state with a large $T^{AM}/T_c^{onset}$. With decreasing $\bar{\gamma}$ from 5.00 to 1.00, the $T^{AM}/T_c^{onset}$ [marked as the crossing points in Fig. 4(d)] decreases. Therefore, according to the theoretical simulation, the $T^{AM}/T_c^{onset}$ is positively correlated to the $\bar{\gamma}$ and hence shows a negative relation to the average link resistance $\overline{R_L}$ and the normal state resistance $R_N$ consistent with our experimental observations [Fig. 4(c)].

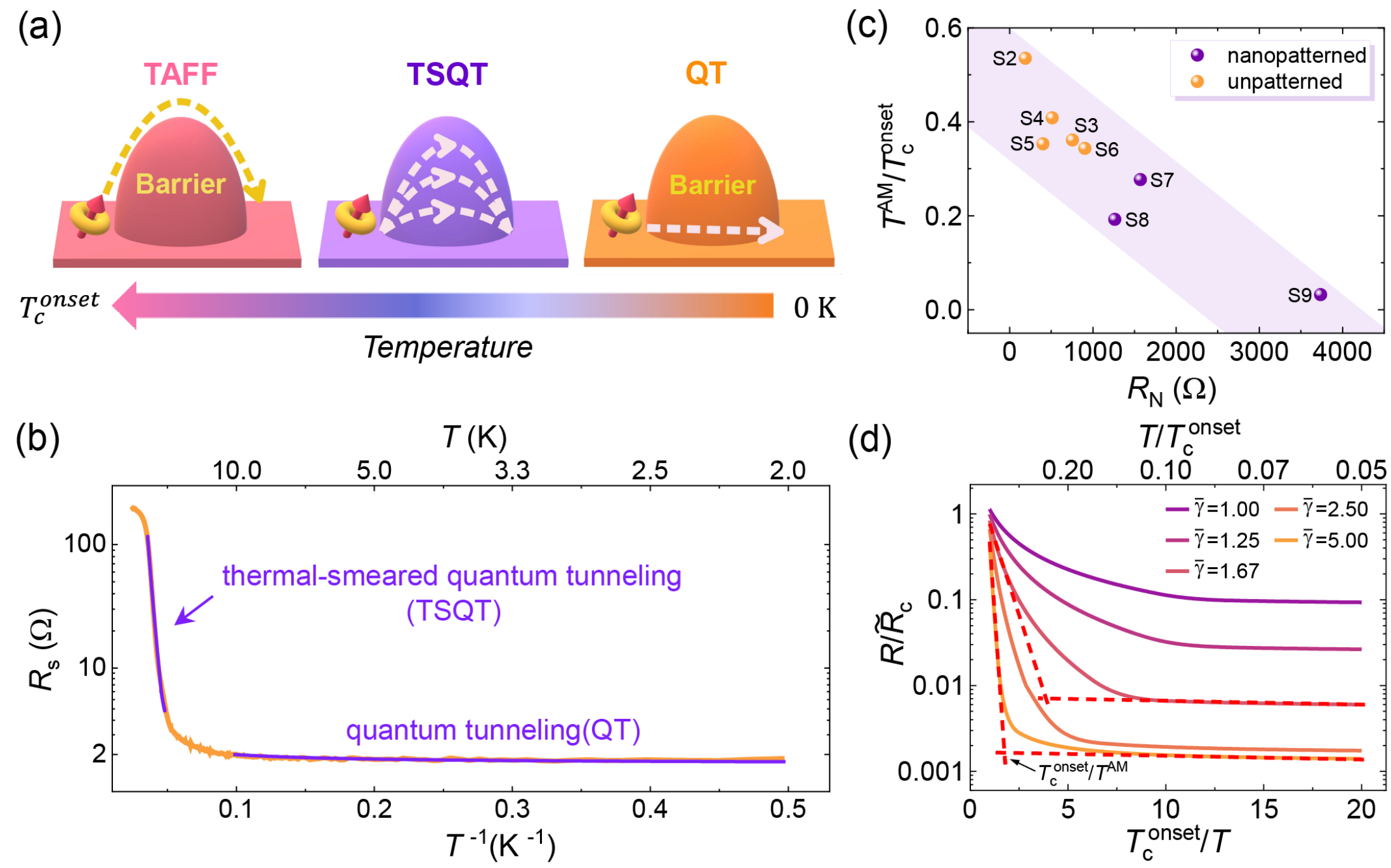

FIG. 4. The origin and evolution of anomalous metal states. (a) Schematic diagram of vortex dynamics. Thermal activated flux flow (TAFF) is dominant at relatively high temperatures where the vortex motion is driven by thermal fluctuations. Quantum tunneling (QT) of vortices is dominant at low temperatures where the vortex motion is driven by quantum fluctuations. The thermal smeared quantum tunneling (TSQT) becomes dominant in the moderate temperature regime. (b) The $\mathrm{lg}R_s$ versus $1/T$ curve of a high-temperature anomalous metal state sample ($S2$, orange line)) and the fitting results (purple lines) based on our theoretical model considering quantum tunneling of vortices in the low temperature regime and thermal-smeared quantum tunneling in the moderate temperature regime. The corresponding temperature $T$ is labeled at the top axis. (c) $T^{\mathrm{AM}}/T_{\mathrm{c}}^{\mathrm{onset}}$ as a function of normal state resistance ($R_{\mathrm{N}}$) at zero field for both unpatterned and nanopatterned films ($S2$ to $S9$). The shadow areas are guides for the eye. (d) Quantitative simulation of the resistance-temperature behavior from quantum tunneling of vortices. The panel shows $\mathrm{lg}\, R/\widetilde{R_{\mathrm{c}}}$ versus $T_{\mathrm{c}}^{\mathrm{onset}}/T$ curves with different colors representing different dissipation strength $\bar{\gamma}$ ($\bar{\gamma} = \frac{h}{4e^2 \overline{R_L}}$ and $\overline{R_L}$ is the average link resistance). The top axis indicates the corresponding $T/T_{\mathrm{c}}^{\mathrm{onset}}$ value. Here, we assume $T_{\mathrm{c}}^{\mathrm{onset}} = 50\ K$ and $E_{\mathrm{J}} = \pi k_{\mathrm{B}} T_{\mathrm{c}}^{\mathrm{onset}}$ and set $\widetilde{R_{\mathrm{c}}} = \frac{h}{4e^2} \cdot \frac{E_{\mathrm{c}}}{4 k_{\mathrm{B}} T_{\mathrm{c}}^{\mathrm{onset}}} \cdot \frac{\sqrt{\pi}}{2}$. The resistance curves have two parts (i.e., thermal-smeared quantum tunneling and quantum tunneling of vortices) connected by a smooth Bezier function. The value of $T^{\mathrm{AM}}/T_{\mathrm{c}}^{\mathrm{onset}}$ can be calculated from the crossing point of the red dashed lines.

In summary, we report the observation of a high-temperature anomalous metal state in crystalline FeSe/STO interface superconducting systems. The exceptionally high $T^{\mathrm{AM}}$ and $T^{\mathrm{AM}}/T_{\mathrm{c}}^{\mathrm{onset}}$ of the anomalous metal state observed under zero magnetic field results from the quantum tunneling of vortices with the strong Ohmic dissipation. We demonstrate that the $T^{\mathrm{AM}}$ can be changed by orders of magnitude via the fabrication of nano-hole arrays, revealing the effective modulation of anomalous metal states. Our investigation enriches the understanding of the anomalous metal states, and the significantly high $T^{\mathrm{AM}}$ could bring in other experimental techniques (e.g., infrared optical conductivity), which are very difficult to be carried out at ultralow temperatures, on the studies of anomalous metal states. Moreover, further understanding of quantum dynamics of vortices calls for systematic investigations on the scaling properties between the Hall resistance and the longitudinal resistance for the anomalous metal states in both unpatterned and nanopatterned 2D superconducting systems.

## ACKNOWLEDGMENTS

We thank Xucun Ma, Guanyang He, Jie Liu, Rao Li, and Yue Tang for their help in sample fabrication and characterization. We thank Zheyuan Xie for assistance in theoretical fitting. We thank Yanzhao Liu, Ying Xing, Ziqiao Wang and Jun Ge for helpful discussions. We thank Sarah Sun and Quantum Design China for the help in the transport measurement. This work was financially supported by the National Natural Science Foundation of China (Grant No. 12488201), the National Key Research and Development Program of China (Grant No. 2022YFA1403103, No. 2018YFA0305604), the National Natural Science Foundation of China (No. 11774008, No. 51788104, No. 12074210, No. 12174442 , No. 12374037), the Innovation Program for Quantum Science and Technology (2021ZD0302403), Beijing Natural Science Foundation (Z180010), the Basic and Applied Basic Research Major Programme of Guangdong Province, China (Grant No. 2021B0301030003), Jihua Laboratory (Project No. X210141TL210), Beijing National Laboratory for Condensed Matter Physics, Young Elite Scientists Sponsorship Program by BAST (No. BYESS2023452), the Fundamental Research Funds for the Central Universities and the Research Funds of Renmin University of China (Grant No. 22XNKJ20).

Yanan Li, H. L. and H. J. contributed equally to this work.

# Supplemental Materials for

# High-Temperature Anomalous Metal States in Iron-Based Interface Superconductors

Yanan Li[1,6], Haiwen Liu[2], Haoran Ji[1], Chengcheng Ji[1,9], Shichao Qi[1], Xiaotong Jiao[3,5], Wenfeng Dong[3], Yi Sun[1], Wenhao Zhang[3,8], Zihan Cui[4], Minghu Pan[5], Nitin Samarth[6], Lili Wang[3], X. C. Xie[1,9,10,11], Qi-Kun Xue[3,9,12], Yi Liu[4,7*], and Jian Wang[1,9,10‡]

[1]*International Center for Quantum Materials, School of Physics, Peking University, Beijing 100871, China*

[2]*Center for Advanced Quantum Studies, Department of Physics, Beijing Normal University, Beijing 100875, China*

[3]*State Key Laboratory of Low-Dimensional Quantum Physics, Department of Physics, Tsinghua University, Beijing 100084, China*

[4]*Department of Physics and Beijing Key Laboratory of Opto-electronic Functional Materials & Micro-nano Devices, Renmin University of China, Beijing 100872, China*

[5]*School of Physics and Information Technology, Shaanxi Normal University, Xi'an 710119, China*

[6]*Department of Physics, The Pennsylvania State University, University Park, Pennsylvania 16802, USA*

[7]*Key Laboratory of Quantum State Construction and Manipulation (Ministry of Education), Renmin University of China, Beijing 100872, China*

[8]*School of Physics and Wuhan National High Magnetic Field Center, Huazhong University of Science and Technology, Wuhan 430074, China*

[9]*Hefei National Laboratory, Hefei 230088, China*

[10]*Collaborative Innovation Center of Quantum Matter, Beijing 100871, China*

[11]*Institute for Nanoelectronic Devices and Quantum Computing, Fudan University, Shanghai 200433, China*

[12]*Southern University of Science and Technology, Shenzhen 518055, China*

Yanan Li, H. L. and H.J. contributed equally to this work.

‡Corresponding author. jianwangphysics@pku.edu.cn
*Corresponding author. yiliu@ruc.edu.cn

## I. Methods

**Film growth.** Thin FeSe films (1-5 unit-cells) were epitaxially grown in an ultra-high vacuum molecular beam epitaxy (MBE) chamber. Pretreated single crystal STO (001) substrates were used for the growth[1]. We etched the substrates with deionized water (90 ℃, 45 min) and 10% HCl solution (room temperature, 45 min). Then, we annealed the substrates in a tube furnace under oxygen flow at 980 ℃ for 3 h. Before growth, the substrates were degassed at 600 ℃ for 0.5 h in MBE chamber. Through the above treatments, the STO (001) surface became atomically flat with step terrace structure and $TiO_2$ termination. FeSe films were then grown by co-evaporating Fe (99.995%) and Se (99.9999%) from Knudsen cells with a flux ratio of ~1:10 as the substrates were heated to 400 ℃. After growth, FeSe films were gradually annealed up to 450 ℃. To protect the thin FeSe films from oxidization, FeTe protection layers were grown by co-evaporating Fe (99.995%) and Te (99.9999%) with a flux ratio of ~1:4 at 270 ℃. During the growth, scanning tunneling microscope was used to examine the morphology and crystal quality of the films. Our films are rectangular strips around 6 mm long and 2 mm wide, grown on substrates in 2 mm × 10 mm.

**The fabrication of nano-holes array.** To fabricate Josephson junction array (JJA) on FeSe/STO, we etched the FeSe films by reactive ion etching (RIE) technique through anodic aluminum oxide (AAO) membrane masks. The AAO mask, with a triangular array of holes (70 nm in diameter and 100 nm in period), was transferred onto the FeSe film in acetone[2]. The etching was performed with Ar flow (20 sccm) and 200 W radio frequency power. The chamber pressure was kept around 6.0 Pa during etching. With increasing etching time, the normal state resistance increases and the film changes from superconducting state to anomalous metal and then insulating state.

**Transport measurements.** The six-probe configuration was used for the transport measurements, as shown in the inset of Fig. 1(a). Two indium strips ($I$+ and $I$-) were pressed along the width of the film so that the current could homogeneously pass through the sample. The other four indium electrodes acted as the voltage probes. Two of them ($V$+ and $V$-) were used to measure the longitudinal voltage and another two electrodes ($V_H$) were Hall electrodes. We use a small excitation current within the linear regime of the $I$-$V$ curve for all the resistance measurements in this paper unless stated otherwise. The temperature-dependent resistance and magnetoresistance were measured in a Physical Property Measurement System (*Quantum Design*). Ultra-low temperature measurements down to 50 mK were carried out in a dilution refrigerator option with radio frequency filters (*Quantum Design*).

## II. Quantum oscillations in nanopatterned FeSe/STO

The $h/2e$ oscillations are observed in both the anomalous metal state and the insulating state (Fig. S10 and Fig. S14), which demonstrates that Cooper pairs dominate the transport in these states. The temperature dependence of the oscillation amplitude is extracted as $G_{\mathrm{osc}} = \left| G\left(\frac{B_0}{2}\right) - G_0 \right|$ after subtracting the background, where $G_0$ and $G\left(\frac{B_0}{2}\right)$ are the conductance of peak and dip of each oscillation, as shown in Fig. S10(d). The phase coherence length ($L_\phi$) of Cooper pairs is estimated by $G_{\mathrm{osc}} = \frac{4e^2}{h}\left(\frac{L_\phi}{\pi r}\right)^{1.5}\exp\left(-\frac{\pi r L_\phi}{L_\phi}\right)$[2]. r is half of the center-to-center hole spacing ~51 nm. $G_{\mathrm{Q}}$ is the quantum conductance for Cooper pairs, $\frac{4e^2}{h}$. $G_{\mathrm{osc}}$ and $L_\phi$ saturate at low temperatures (Fig. S10(d)), reminiscent of the anomalous metal state in nanopatterned $YBa_2Cu_3O_{7-x}$ films[2].

For the samples in our investigation (the diameter of hole is comparable to the center-to-center hole spacing), the period of quantum oscillations does not depend on the hole sizes, but depend on the area of one unit cell of the nanopattern (i.e., nano-hole array). If the diameter of hole is much smaller than the hole spacing, presumably both these parameters may influence the period of quantum oscillation, which is not the case for our current samples and is beyond the scope of this work.

## III. Bosonic strange metal state in FeSe/STO

In the wide superconducting transition region of FeSe films with a larger normal state resistance, we observe an extraordinary linear-in-temperature ($T$-linear) resistance below $T_{\mathrm{c}}^{\mathrm{onset}}$. Figure S3 summarizes the $T$-linear resistance under perpendicular magnetic fields. The $T$-linear resistance extends to lower temperatures (Figs. S3(a) and S3(d)), and the corresponding temperature regime ($\Delta T/T_{\mathrm{c}}^{\mathrm{onset}}$) grows with increasing magnetic field (Figs. S3(b) and S3(e)) when the field is relatively small. Under a higher magnetic field, $T_{\mathrm{c}}^{\mathrm{onset}}$ decreases and the temperature regime of $T$-linear resistance shrinks. This $T$-linear resistance behavior is reminiscent of the strange metal state in fermionic strongly correlated systems such as the cuprates[3-5], pnictides[6,7], heavy fermion systems[8,9], and magic angle graphene[10]. The fermionic strange metal is associated with the quantum criticality of unconventional superconductivity or magnetism, where pseudogap[11], nematic order[12], and ferromagnetic[13] or antiferromagnetic[14] order are suppressed. In our measurements, the slopes $\alpha_{\mathrm{B}}$ of $T$-linear resistance below $T_{\mathrm{c}}^{\mathrm{onset}}$ are summarized in Figs. S3(c) and S3(f). With increasing magnetic field, $\alpha_{\mathrm{B}}$ firstly decreases and then increases, while $\Delta T/T_{\mathrm{c}}^{\mathrm{onset}}$ firstly increases and then decreases (Figs. S3(b) and S3(e)). At present, the theoretical model of bosonic metal states in this work does not consider the case of finite magnetic field and thus cannot explain the non-monotonic variation of $\alpha_{\mathrm{B}}$. Moreover, at zero field, the slope of the FeSe film is around 34-120 Ω/K, a larger value compared with previous observations of fermionic strange metals. To be specific, the slope $\alpha_{\mathrm{F}}$ for high-temperature cuprate superconductors (e.g. $La_{2-x}Sr_xCuO_4$, $Pr_{2-x}Ce_xCuO_{4\pm\delta}$, and $La_{2-x}Ce_xCuO_4$) lies in the range 1.7-8.2 Ω/K[4]. Furthermore, recent studies report that $\alpha_{\mathrm{F}}$ is even smaller (~0.13 Ω/K) in thick FeSe films grown by pulsed-laser deposition [15]. In our measurements, above $T_{\mathrm{c}}^{\mathrm{onset}}$ one typical pristine 1-UC FeSe/STO shows $T$-linear resistance with the slope of 1.23 Ω/K (Fig. S2(b)). Therefore, the large slope observed in our FeSe films below $T_{\mathrm{c}}^{\mathrm{onset}}$ is distinct from the fermionic strange metal behavior reported in the previous works. Additionally, the Hall coefficient is significantly suppressed with decreasing temperature in the temperature regime of $T$-linear resistance (Figs. S6(c) and S6(d)). Thus, the strange metal state below $T_{\mathrm{c}}^{\mathrm{onset}}$ mostly originates from the transport of Cooper pairs (bosons) rather than quasiparticles (fermions).

Furthermore, we would like to emphasize that, although the transport properties of anomalous metal and bosonic strange metal states are dominated by Cooper pairs, the influence of fermionic quasiparticles is necessary. Theoretically, in a JJA system, the electric resistance of a bosonic system can be attributed to the motion or quantum tunneling of vortices associated with the Josephson junction[16]. The coupling between bosonic modes and fermionic modes leads to the Ohmic dissipation[17], which could change the dynamics for the quantum tunneling of vortices in the JJA, and gives rise to the anomalous metal and bosonic strange metal states. Therefore, the anomalous metal and bosonic strange metal states are dominated by Cooper pairs, but the quantum tunneling of vortices in these bosonic quantum states are influenced by the dissipation, originating from the couplings to the fermionic quasiparticles.

For the FeSe films with triangular array of nano-holes in Fig. S4(a), the 0 s and 90 s etched films exhibit a zero-resistance state at low temperatures. The superconducting transition region gets wider as $R_{\mathrm{N}}$ increases. Meanwhile, below $T_c^{\mathrm{onset}}$ a remarkably $T$-linear resistance temperature range (Fig. S4(a)) appears and extends with increasing etching time from 90 s ($R_{\mathrm{N}} = 1.2$ kΩ) to 210 s ($R_{\mathrm{N}} = 3.7$ kΩ). The film with the etching time of 270 s shows insulating behavior at low temperatures ($R_{\mathrm{N}}$ is as high as 31.5 kΩ). Figure S5 presents the temperature dependence of $R_s$ and $dR_s/dT$ for the 150 s and 210 s etched FeSe/STO samples showing bosonic strange metal state. The $dR_s/dT$ vs $T$ curve approximately shows a plateau from 1.8 K to 7.1 K (1.2 K to 8.1 K) for 150s (210s) etched sample (as denoted by the shadow areas), confirming linear temperature dependence of the sheet resistance.

To further demonstrate the $T$-linear resistance, we also fit the $R_{\mathrm{s}}(T)$ curves of FeSe/STO samples with different etching time or at various magnetic fields by using the power law equation $R_{\mathrm{s}}(T) = a + bT^n$, where $a$, $b$ and $n$ are fitting parameters. As shown in Table S2, $n$ is around 1 for most $R_{\mathrm{s}}(T)$ curves, verifying the $T$-linear resistance. Under a higher magnetic field, $T_{\mathrm{c}}^{\mathrm{onset}}$ decreases and the temperature regime of strange metal state shrinks with $n$ deviating from 1.

**IV. Exclusion of classical percolation model for anomalous metal state**

The anomalous metal state in FeSe/STO interface superconductor exhibits various interesting phenomena, including resistance saturation at low temperatures, zero Hall resistance, large magnetoresistance and linear *I-V* curves. These features cannot be explained by the classical percolation model[18], where the local superconducting regions are embedded in the normal state. In such case, although the combination of the zero-resistance superconducting and normal state regions may satisfy the requirement of linear *I-V* curve, large magnetoresistance is not expected (the magnetoresistance is very small for the normal state as presented in Fig. S15), which are inconsistent with our observations. In addition, if the finite resistance of the anomalous metal state is contributed by the normal state, the Hall resistance of the system should be non-zero, contradictory to our experimental observation[19]. Therefore, the above discussion demonstrates that the classical percolation model cannot explain the anomalous metal state.

As shown by previous scanning tunneling microscopy (STM) studies, the atomically flat FeSe films contains different domains[20]. The domain walls may behave like Josephson weak links in the FeSe samples. When the applied current exceeds the Josephson critical current of the weak links, the I-V curve of the sample would be non-linear. However, in our measurement, we observed linear I-V curves (Fig. S4(c), inset of Fig. 1(b) and Fig. S7) for the anomalous metal and bosonic strange metal states starting from zero current. Therefore, the observed residual resistance at low temperatures cannot be explained by the assumption that the current in the measurements exceeds the Josephson critical current. Moreover, the observation of zero-Hall coefficient in the anomalous metal state reveals the particle-hole symmetry from Cooper pairs and further excludes the possibility that the residual resistance is from the normal state between the superconducting domains.

**V. Discussion about the influence from upper FeSe layers and FeTe capping layer on the bosonic metal states**

In pristine FeTe/STO, the sheet resistance is above 3700 Ω below 20 K[1]. In this work, the resistance of anomalous metal states saturates to a small finite value (around 0.1-10 Ω) in pristine FeSe/STO, around 10-100 Ω in nanopatterned FeSe/STO at low temperatures (Fig. S11 and S12). The saturated resistance value of anomalous metal state in both pristine and nanopatterned FeSe/STO is far smaller than the resistance of pristine FeTe film at low temperatures. Therefore, the influence of FeTe capping layer on the anomalous metal behavior can be negligible.

As for the possible influence of FeTe capping layer on the bosonic strange metal state, control experiments have been performed to study the effect of FeTe layer on nanopatterned FeSe/STO. The FeTe/STO system is grown and then etched for 270 s with the same procedure as FeSe/STO samples. The nanopatterned FeTe/STO system becomes very insulating with the sheet resistance as large as $2 \times 10^7$ Ω at 300 K, several orders of magnitude larger than that of the 270 s etched FeSe/STO. Therefore, the resistance of nanopatterned FeTe capping layer should be much larger than that of the corresponding FeSe/STO samples with the same etching time. Thus, the transport properties in the nanopatterned FeSe/STO cannot be attributed to the FeTe capping layer.

Furthermore, in our current theoretical consideration, the coupling between the bosonic modes and the fermionic modes contributes to the Ohmic dissipation[17] in the two-dimensional (2D) dissipative quantum XY model with the dissipation coefficient $\gamma = \frac{h}{4e^2 R_{\mathrm{L}}}$ (here the resistance $R_{\mathrm{L}}$ collects the contribution of the fermionic modes). Thus, the possible influence of upper FeSe layers and FeTe capping layer as the parallel fermionic channels has been considered in the dissipation coefficient $\gamma$.

The previous work[21] reports that a parallel metallic ground plane prevents the formation of anomalous metal states in 2D superconducting systems. It is proposed that the capacitive coupling of the metallic ground plane causes the suppression of the quantum fluctuations and helps restore the superconducting state. The FeTe capping layer on our ultrathin FeSe films is not metallic but semiconducting[1]. Thus, the FeTe capping layer cannot be regarded as the metallic ground plane at low temperatures. The FeTe capping layer may change the link resistance and thus influence the dissipation strength of the system, which has been considered in the dissipation coefficient $\gamma$ in our model. In our work, we find that for a nonuniform system a moderate dissipation strength $\gamma$ around 1 contributes to the high temperate bosonic strange metal state, and the Ohmic dissipation strength $\gamma$ within the range [0.5, 1] contributes to the zero-temperature anomalous metal state. If Ohmic dissipation strength $\gamma > 1$ at zero temperature or $\gamma \gg 1$ at finite temperatures, the Ohmic dissipation damps the phase fluctuation and stabilizes the superconducting state, similar to the results of previous work[21].

**VI. Discussion on the origin of anomalous metal and bosonic strange metal states**

The electric resistance in a pure bosonic system can be attributed to the classical motion or quantum tunneling of vortices[22]. The well-known examples of classical motion include case A the free vortices motion above BKT temperature[23], and case B the thermal activated flux flow under magnetic field[24]. The above-mentioned case A and case B has one respective difference that in case A the resistance is mainly dominated by the proliferation of vortex density while for case B the resistance is mainly determined by the diffusion coefficient. Without dissipation, when the temperature decreases to approaching zero, the quantum tunneling events of vortices become predominant (resulting in an insulator) or reduce to zero (resulting in a superconductor). This novel type of superconductor-insulator transition (SIT) can be attributed to competition between the charging energy $E_c$ and Josephson energy $E_J$, and this SIT usually possesses the characteristic of duality between charge and vortices with the critical quantum resistance $R_c = \frac{h}{4e^2}$ [25,26]. On the other hand, the coupling between fermionic modes and bosonic modes leads to the Ohmic dissipation[27], and the Ohmic dissipation effect will change the microscopic dynamics of vortices[28-30], which further gives rise to observable consequences in experiments[31-33]. In the following, we demonstrate that the Ohmic dissipation with moderate strength fully changes the dynamics of vortices tunneling beyond the aforementioned two cases and gives rise to the experimental features of anomalous metal and bosonic strange metal.

The 2D dissipative quantum XY model provides a microscopic basis for the analysis of vortices dynamics[17,30,34] with the action:

$$S = \sum_j \int d\tau \left\{ \frac{|\dot{\theta}_j|^2}{2E_{c,j}} - \sum_{<i,j>} E_J Cos[\theta_i - \theta_k] \right\} + \sum_j \frac{\gamma_j}{4\pi} \int d\tau_1 d\tau_2 \left| \frac{e^{i\theta_j(\tau_1)} - e^{i\theta_j(\tau_2)}}{\tau_1 - \tau_2} \right|^2 . \quad \text{(SE1)}$$

Here, $E_{c,j}$ denotes the charge energy at site j, $E_J$ denotes the local Josephson energy and $\gamma_j = \frac{h}{4e^2 R_L}$ denotes the dimensionless dissipation strength determined by the link resistance $R_L$. The parameters $E_c$, $E_J$ and $\gamma_j$ are in general random numbers.

(A) The tunneling rate of vortices in a uniform system

Previously, in systems with homogenous $\gamma$, it is shown that the dissipation can drive a quantum phase transition between superconductor and metal[31]. In order to combine this zero-temperature phase diagram with quantum dynamics of vortices, we firstly focus on the action at one single Josephson junction, namely the shunted Josephson junction model[31,35]:

$$S = \int d\tau \left\{ \frac{|\dot{\theta}|^2}{2E_c} - E_J Cos[\theta] \right\} + \frac{\gamma}{4\pi} \int d\tau_1 d\tau_2 \left| \frac{\theta(\tau_1) - \theta(\tau_2)}{\tau_1 - \tau_2} \right|^2 . \quad \text{(SE2)}$$

This action can be viewed as a dissipative quantum diffusion phase in the potential $E_J Cos[\theta]$. When we discuss the dissipation driven SIT, we consider the case $E_c \ll E_J$ and the $\gamma$ is of order one. Thus, for 2D dissipative quantum XY model, the two representative energy scale is the high frequency cutoff $\omega_c = \frac{E_J}{\hbar}$ and the low-energy plasma frequency $\omega_p = \frac{\sqrt{E_c \cdot E_J}}{\hbar}$ [36]. The dynamics of the phase $\theta$ can be mapped to the dynamics of the spin population $n_s(t)$ in the spin-boson model $\langle n_s(t) n_s(0) \rangle \propto e^{\frac{-t}{\tau_s}}$ with the tunneling rate of vortices in the moderate temperature regime obeying the relation[37]:

$$\frac{1}{\tau_s} = \frac{\omega_\mathrm{p}^2}{\omega_\mathrm{c}} \frac{\sqrt{\pi}}{2} \frac{\Gamma(\gamma)}{\Gamma(\gamma + 1/2)} \left( \frac{\pi k_B T}{\hbar \omega_c} \right)^{2\gamma - 1} . \quad \text{(SE3)}$$

One can directly find around the critical link resistance $R_L = \frac{h}{4e^2}$ with $\gamma = 1$, the tunneling rate of vortices has a very simple form:

$$\frac{1}{\tau_s} = \frac{E_C}{E_J}\frac{\pi k_B T}{\hbar}. \text{ (SE4)}$$

We need to mention that around the critical link resistance the tunneling rate of vortices generally satisfies the relation $\frac{\hbar}{\tau_s} = O(1) \cdot k_B T$. Considering the above tunneling rate of vortices with the resistance in the Drude form, one can obtain the sheet resistance:

$$R_s = \frac{m^*}{n_s(2e)^2} \cdot \frac{1}{\tau_s}. \text{ (SE5)}$$

And considering the relation between Cooper pair mass $m^*$ and superfluid density $n_s$ and the 2D superconducting (onset) temperature $T_c^{\text{onset}}$ with relation $\frac{\pi\hbar^2 n_s}{2m^*} = k_B T_c^{\text{onset}}$ [38,39], the resistance reads:

$$R_s = \frac{h\hbar}{4k_B T_c^{\text{onset}}(2e)^2} \cdot \frac{1}{\tau_s}. \text{ (SE6)}$$

Thus, the general resistance can be obtained $R_s \propto T^{2\gamma-1}$. Around the critical link resistance with $\gamma = 1$, the resistance has a linear *R-T* form:

$$R_s = \frac{\pi E_C}{4E_J}\frac{h}{(2e)^2} \cdot \frac{T}{T_c^{\text{onset}}}, \text{ (SE7)}$$

with the slope of *T*-linear resistance $dR_s/dT = \frac{\pi E_C}{4E_J}\frac{h}{(2e)^2 T_c^{\text{onset}}}$. The optimal slope of linear *R-T* is approaching $\frac{h}{(2e)^2 T_c^{\text{onset}}}$, but the general case of *R-T* is smaller due to the factor $\frac{\pi E_C}{4E_J}$. Based on the linear *R-T* form $R_s = \frac{\pi E_C}{4E_J}\frac{h}{(2e)^2} \cdot \frac{T}{T_c^{\text{onset}}}$ around $\gamma = 1$, the slope of the system with similar $T_c^{\text{onset}}$ is predominantly determined by $\frac{E_C}{E_J}$, and mainly represents the tunneling rate of phase difference $\theta$ at the Josephson junction with the form $\frac{1}{\tau_s} = \frac{E_C}{E_J}\frac{\pi k_B T}{\hbar}$ for the system with given $T_c^{\text{onset}}$. In the samples with smaller $R_N$, the superconducting islands are larger with relatively smaller $E_C$ and smaller $\frac{E_C}{E_J}$, thus the slope positively correlates with $R_N$ (Fig. S17).

The Landau overdamped (z=2) mechanism on the pairing field also provides a phenomenological explanation for the marginal Fermi liquid[40-42]. Here, we mainly focus on the contribution from the microscopic vortices dynamics on the measured resistance in a 2D system to understand both the bosonic anomalous metal and strange metal states, thus we start from the 2D dissipative XY model as shown in Eq. (SE1).

(B) The diffusion of vortices in a uniform system

The tunneling rate of vortices can also link with the diffusion coefficient of vortices. In the following, we will show that both the spatial-domain diffusion process of vortices (shown in Eq. (SE9)) and the time-domain tunneling process of phase (shown in Eq. (SE3)) give the same result to illustrate the resistance of 2D dissipative superconductors. Generally, the resistance in a 2D superconductor thin film originates from the mobile vortices with the novel relation[23]:

$$R_s = \frac{h^2}{4e^2} n_v \mu_v, \text{ (SE8)}$$

here $n_v$ denotes the density of vortices and the mobility of vortices satisfies the relation $\mu_v = \frac{\tau_v}{m_v}$ with $\tau_v$ and $m_v$ denotes the mean free time and mass of vortices, respectively. Considering the Drude form of resistance in Eq. (SE5), one can rewrite Eq. (SE8) in the following convenient form:

$$n_s \mu_s n_v \mu_v \approx h^2. \text{ (SE9)}$$

In the following, we discuss this relation in the moderate temperature regime and in the zero-temperature regime in the 2D dissipative quantum XY model. The diffusion of vortices in this system can be viewed as the quantum Brownian motion vortices in corrugated potential with dissipation, and the mobility of the systems in the moderate temperature regime reads[29]:

$$\mu_v = \mu_{v,0} \cdot \frac{\sqrt{\pi}}{2} \frac{\Gamma(\gamma)}{\Gamma(\gamma+1/2)} \left(\frac{\pi k_B T}{\hbar \omega_c}\right)^{2\gamma-1}, \text{(SE10)}$$

here $\mu_{v,0} \approx A\frac{d^2}{h}$ denotes the bare vortices mobility with the inter-vortices distance $d$ and a prefactor A. One can easily find that the mobility of vortices in Eq. (SE10) decreases with decreasing temperature for $\gamma > \frac{1}{2}$ and acts oppositely for $\gamma < \frac{1}{2}$. This behavior indicates a quantum phase transition around $\gamma = \frac{1}{2}$, which is consistent with the previous prediction in 2D dissipative superconductors with $E_c \ll E_J$ [36]. We define the characteristic time of Cooper pair $\frac{1}{\tau_{s,0}} = \frac{\omega_{\mathrm{p}}^2}{\omega_{\mathrm{c}}} = \frac{E_c}{\hbar}$, and $\mu_{s,0} = \frac{\tau_{s,0}}{m^*}$. Thus, the bare parameters with relation $n_s \mu_{s,0} n_v \mu_{v,0} \approx h^2$ gives a constraint for the perfector $\mathrm{A} = \frac{E_c}{4 k_B T_{\mathrm{c}}^{\mathrm{onset}}}$. Then, with the moderate temperature vortices mobility relation (SE10) and the moderate temperature tunneling rate of vortices in the (SE3), one can check equation (SE9) is valid for the general case.

We want to elaborate on the importance of Eq. (SE9) and briefly mention why this relation can be generalized beyond the well-known examples of resistance in superconductors including the free vortices motion in BKT transition and the thermal activated flux flow. Previously, without dissipation, the quantum phase transition of SIT driven by the competition between charge energy and Josephson energy is well established[25,26]. Meanwhile, a strong dissipation with $\gamma \gg 1$ and $E_c \ll E_J$ will stabilize the superconducting phase; no quantum phase transition but thermal BKT transition occurs. Thus, under influence of dissipation, interesting quantum dynamics of vortices emerges in the moderate strength case with $\gamma \sim O(1)$. The spatial-domain diffusion analysis of vortices applies both for the uniform systems and disordered systems.

(C) The resistance in the zero-temperature limit and the influence of randomness in $\gamma$ on the resistance

The tunneling rate and diffusion coefficient in Eq. (SE3) and (SE10) break down in the zero-temperature limit[28,29,37]. In the zero-temperature limit, the renormalized mobility of vortices manifests three respective features[29]: for $\gamma > 1$ the vortices localizes with $\mu_v = 0$, for $\gamma < \frac{1}{2}$ the dynamics of vortices cannot be viewed as a diffusion process, and for $\frac{1}{2} < \gamma < 1$, the mobility reads:

$$\mu_v = \frac{C}{2\sqrt{\pi}} \cdot \mu_{v,0} \cdot \frac{\Gamma\left(\frac{\gamma}{2(1-\gamma)}\right)}{\Gamma\left(\frac{1}{2(1-\gamma)}\right)}. \text{ (SE11)}$$

The mobility as a function of $\gamma$ in Eq. (SE11) smoothly connects with the zero mobility for $\gamma > 1$. We also note another recent theoretical prediction of nonzero mobility within the regime $\frac{1}{2} < \gamma < 1$ with different functional form of $\gamma$ [36]. This nonzero mobility of vortices will give rise to a resistance in the low-temperature regime:

$$R_s = \frac{h^2}{8\sqrt{\pi} e^2} n_v C \cdot \mu_{v,0} \cdot \frac{\Gamma\left(\frac{\gamma}{2(1-\gamma)}\right)}{\Gamma\left(\frac{1}{2(1-\gamma)}\right)}, \text{(SE12)}$$

here $C$ is a constant of order one, and we choose $C \approx 1$ for simplicity in the following. The low-temperature quantum tunneling of vortices can give rise to the anomalous quantum metal in experiments.

We next discuss the effect of random link resistance $R_L$ and thus random $\gamma_j$. We assume the link resistance $R_L$ satisfies a Gamma distribution $P_\Gamma(R_L)$ with the mean link resistance $\overline{R_L}$ and standard deviation $\sigma_R$. The mean link resistance $\overline{R_L}$ gives the mean value of dimensionless dissipation parameter $\bar{\gamma} = \frac{h}{4e^2\overline{R_L}}$ (the random dissipation parameter $\gamma = \frac{h}{4e^2 R_L}$), and the moderate temperature resistance can be approximated by the simple version contributed from the thermal-smeared quantum tunneling of vortices:

$$R_S = \frac{h}{4e^2}\cdot\frac{E_C}{4k_B T_c^{\text{onset}}}\cdot\frac{\sqrt{\pi}}{2}\frac{\Gamma(\bar{\gamma})}{\Gamma(\bar{\gamma}+1/2)}\left(\frac{\pi k_B T}{E_J}\right)^{2\bar{\gamma}-1}. \text{ (SE13)}$$

Here we utilize the approximation $n_v\mu_{v,0} \approx A n_v \frac{d^2}{h} \approx \frac{A}{h} \approx \frac{E_C}{4k_B T_c^{\text{onset}}}\cdot\frac{1}{h}$, with $E_C$ denoting the charging energy and $T_c^{\text{onset}}$ denoting the onset superconducting temperature. The low temperature resistance is contributed by the link resistance $R_Q < R_L < 2R_Q$ (in other words $\frac{1}{2} < \gamma < 1$). Thus, the resistance in the low temperature regime reads:

$$R_S = \frac{C_1 h}{8\sqrt{\pi}e^2}\frac{E_C}{4k_B T_c^{\text{onset}}}\cdot Exp[T/T_0], \text{ (SE14)}$$

with the constant $C_1 = \int_{R_Q}^{2R_Q} P_\Gamma(R_L)\frac{\Gamma\left(\frac{\gamma}{2(1-\gamma)}\right)}{\Gamma\left(\frac{1}{2(1-\gamma)}\right)} dR_L$, and $T_0$ is an intrinsic energy scale, which is related to $E_\mathrm{J}/\pi k_\mathrm{B}$. Thus, the residual resistance is predominated by the distribution of link resistance, and the total dimensionless resistance can be written in a simplified form:

$$\frac{R_S}{\frac{h\sqrt{\pi}}{8e^2}\frac{E_C}{4k_B T_c^{\text{onset}}}} = \begin{cases} \frac{\Gamma(\bar{\gamma})}{\Gamma(\bar{\gamma}+1/2)}\left(\frac{\pi k_B T}{E_J}\right)^{2\bar{\gamma}-1}, & moderate-T\ regime \\ \frac{C_1}{\pi} Exp[T/T_0], & low-T\ regime \end{cases}. \text{ (SE15)}$$

Thus, given the distribution of link resistance, one can briefly draw the resistance versus temperature based on Eq. (SE15). And the crossing point of these two curves separates the moderate-*T* resistance behavior and the low-*T* bosonic anomalous metal feature. We give out the specific case for the crystalline samples and the nanopatterned samples.

For the crystalline samples, the mean value of link resistance $\overline{R_L}$ is smaller than $R_Q$ thus $\bar{\gamma} > 1$, and meanwhile the link is much more randomized with a large $\sigma_R$. Thus, the moderate temperature resistance-temperature relation becomes superlinear, and the resistance quickly diminishes to approaching zero under $T_c^{\text{onset}}$. For samples with smaller $\overline{R_L}$, the resistance decreases quickly. Moreover, for samples with smaller $\overline{R_L}$, the zero-temperature mobile vortices frequency $\mathrm{P}\left(R_Q < R_L < 2R_Q\right)$ is also smaller. Thus, when increasing the normal state resistance in crystalline samples, one can obtain an enlarged bosonic anomalous metal temperature regime, as shown in Fig. S17.

For the nanopatterned samples, the mean value of link resistance $\overline{R_L}$ can be tuned to approach the optimal value $R_Q$ thus $\bar{\gamma} = 1$, and meanwhile the link is relatively uniform with a small $\sigma_R$. Thus, the moderate temperature resistance-temperature relation becomes linear, namely the bosonic strange metal feature. Moreover, due to the small value of standard deviation $\sigma_R$, the link resistance locates near $R_Q$ and the low-temperature mobility of vortices becomes very small as shown in Eq. (SE15). Thus, in the nanopatterned samples with optimal link resistance $\overline{R_L} \approx R_Q$, the bosonic strange metal feature becomes remarkable while the bosonic anomalous metal feature is not obvious.

Based on Eq. (SE15), the experimental $R_s(T)$ curves of anomalous metal states are fitted by the above theoretical model (Fig. S11 and S12). At low temperatures, the $R_s(T)$ curves are fitted by a simplified formula $R_s = R_0 \cdot Exp(\frac{T}{T_0})$, which accounts for the quantum tunneling behavior of vortices. At moderate temperatures, the $R_s(T)$ curves are fitted by the formula $R_s = A \cdot \left(\frac{T}{T^*}\right)^{2\overline{\gamma}-1}$, which accounts for the thermal-smeared quantum tunneling behavior of vortices. The fitting parameters are listed in Table S1.

**(D) Discussions on the relation between the normal state sheet resistance $R_N$ and the bosonic metal states**

Figure S13 presents the $R_s(T)$ curves of the FeSe/STO samples with smaller normal state sheet resistance $R_N$ than that of sample $S2$ (sample $S2$ shows high-temperature anomalous metal state, and $R_N$ of S2 is around 180 Ω). All the FeSe/STO samples are indeed superconducting when their $R_N$ is smaller than 180 Ω and the dissipation strength are supposed to be stronger than that in sample $S2$. This is consistent with our 2D dissipative JJA model, where the stronger dissipation (smaller $R_N$) would more easily pin down the vortices motion in the JJA, and the superconducting (zero-resistance) states are more favored.

In our theoretical model, we assumed the average link resistance $\overline{R_L}$ is roughly proportional to $R_N$, and the $R_N$ is regarded as an indicator for the average Ohmic dissipation strength, which further influences the vortices dynamics, and consequently determines the ground states of the whole system. In practical, however, the link resistances $R_L$ in a real 2D JJA system have a distribution around $\overline{R_L}$. Thus, the local Ohmic dissipation strength $\gamma$ may vary among different local links in the JJA, since $\gamma \propto \frac{1}{R_L}$. As discussed above, the mobility of vortices is nonzero for the regime of $\frac{1}{2} < \gamma < 1$, which corresponds to the finite resistance arising from quantum tunneling of vortices at zero temperature limit (i.e., anomalous metal state). FeSe/STO samples with similar $\overline{R_L}$ (also similar $R_N$, since $R_N$ is proportional to $\overline{R_L}$) may host different distribution of $R_L$, and may show either superconducting state with zero resistance or anomalous metal state with finite resistance at low temperatures, depending on the spatial probability distribution of the link resistance $R_L$. Meanwhile, since the average Ohmic dissipation strength are comparable for the samples with comparable $R_N$, the resistive behaviors of the superconducting transition (the slope, the temperature range, etc.) should be similar, which might be the cause of $T_{AM}$ being comparable with $T_c^{zero}$. This similarity between $T_{AM}$ and $T_c^{zero}$ in FeSe/STO with comparable $R_N$ supports that $R_N$ could be an indicator for the characteristics of the system. Also, as shown in Fig. 4c in the main text, there is a clear correlation between $R_N$ and $T_{AM}$ / $T_c^{onset}$ of the anomalous metal state (shadow area). Therefore, we would like to emphasize that, although there are slight variations, the $R_N$ is an experimentally accessible indicator to characterize the superconducting and anomalous metal states of the 2D system.

## VII. A summary of the previous theoretical understanding of the bosonic metal states and related phenomena

Firstly, around 1990s, experimental studies demonstrated the resistive behavior in the vortex liquid, originated from the thermal assisted flux flow (TAFF) mechanisms. Then, extensive theoretical works have proposed the nonlinear I-V characteristic in the vortex glass phase, the quasi-linear I-V characteristic in the vortex liquid phase, and the thermal melting transition between these phases, which are mainly dominated by thermal activation or thermal fluctuations. These theoretical advances are well summarized in the review articles[43-46].

Secondly, from 1995, A. Kapitulnik's group continued to observe a bosonic metallic state (now called the anomalous metal state) around zero temperatures in the quantum phase transitions of 2D superconductors[47-50]. Meanwhile, Shimshoni et al. proposed a phenomenological quantum percolation model with zero-temperature dissipation and calculated the resistivity of a single tunneling junction around the superconductor-insulator transition (Eqs. 3 and 5 in the work[51]). This model is often referred as "quantum creep of vortex", and utilized in analyzing the magnetic field induced anomalous metal features in recent experimental literatures[52,53]. However, this phenomenological model cannot give the microscopic origin of the zero-temperature dissipation as well as the anomalous metal states. In 2001, A. Kapitulnik and his collaborators proposed that Ohmic dissipation plays an important role in the quantum phase transition of 2D superconducting systems[33]. Meanwhile, many theoretical endeavors proposed to explain the aforementioned zero-temperature anomalous metallic features. In 1998, M. A. Feigel'man and A. Larkin proposed a zero-temperature metallic state may exist in a 2D Josephson coupled array due to the renormalization of Ohmic dissipation strength[54], and then S. Spivak and collaborators promoted this scenario and summarized in the recent review[18].

On the other hand, in 2007, C. M. Varma also considered the important role of Ohmic dissipation in a zero-temperature characteristic of 2D Josephson coupled array[30]. C. M. Varma and collaborators found the renormalization of Ohmic dissipation strength along the renormalization flow can be considered as the "effective screening from instanton and anti-instanton (C. M. Varma and coauthors called them warps)" effect from quantum fluctuation. This is actually reminiscent of the renormalization of superfluid stiffness in BKT transition due to the "effective screening from vortex and anti-vortex" from thermal fluctuation. The theoretical endeavors of C. M. Varma and collaborators are summarized in the review articles[34,55]. Other theoretical proposals for the zero-temperature bosonic metal are provided by P. Philips's group[56] and S. Doniach's group[57]. However, none of these models were able to give the quantitative formula to explain the temperature dependence of resistance in anomalous metal state and couldn't explicitly demonstrate the microscopic mechanism of this phenomenon, especially under zero field. Moreover, the non-zero temperature bosonic strange metal features obviously cannot be explained by the previous theories such as the TAFF mechanism or the BKT transition.

In our work, we closely follow C. M. Varma's approach, but focus on the quantitative explanations of experimental observables, such as the relation between sheet resistance and temperature in the dissipative quantum XY model. Moreover, the aforementioned theoretical works mainly focused on the zero-temperature bosonic metal state, but our theoretical model reveals that both the zero-temperature bosonic anomalous metal state and non-zero temperature bosonic strange metal state originate from the effect of Ohmic dissipation. This consideration seems to have profound influence, explaining the zero-temperature bosonic anomalous metal (which destroys the 2D superfluidity) and the non-zero temperature bosonic strange metal (which destroys the BKT transition) on equal footing of Ohmic dissipation. The microscopic origin for these dramatic influences of Ohmic dissipation is the proliferation of decoupled instanton/anti-instanton (i.e. quantum tunneling of vortices under the influence of Ohmic dissipation), which reflect the mobility and diffusion constant of unpaired vortices. These instanton and anti-instanton are time-domain topological defects generated due to quantum fluctuation in our case, and the spatial-domain topological defects (vortex and anti-vortex) are generated due to thermal fluctuation in the case of BKT transition. In other words, the vortex/anti-vortex decouples into mobile vortex (anti-vortex) when temperature is higher than $T_{\mathrm{BKT}}$, and instanton/anti-instanton decouples into mobile ones when Ohmic dissipation strength $\gamma$<1 at zero temperature or $\gamma$ is smaller than or around 1 at finite temperatures. In our theoretical analysis, we carefully solve the influence of the instanton/anti-instanton on the sheet resistance (as shown in Supplementary Text 4 of the manuscript). Our results firstly give quantitative explanations for both the zero-temperature bosonic anomalous metal and non-zero temperature bosonic strange metal features based on 2D quantum XY model.

## VIII. Figures and Tables

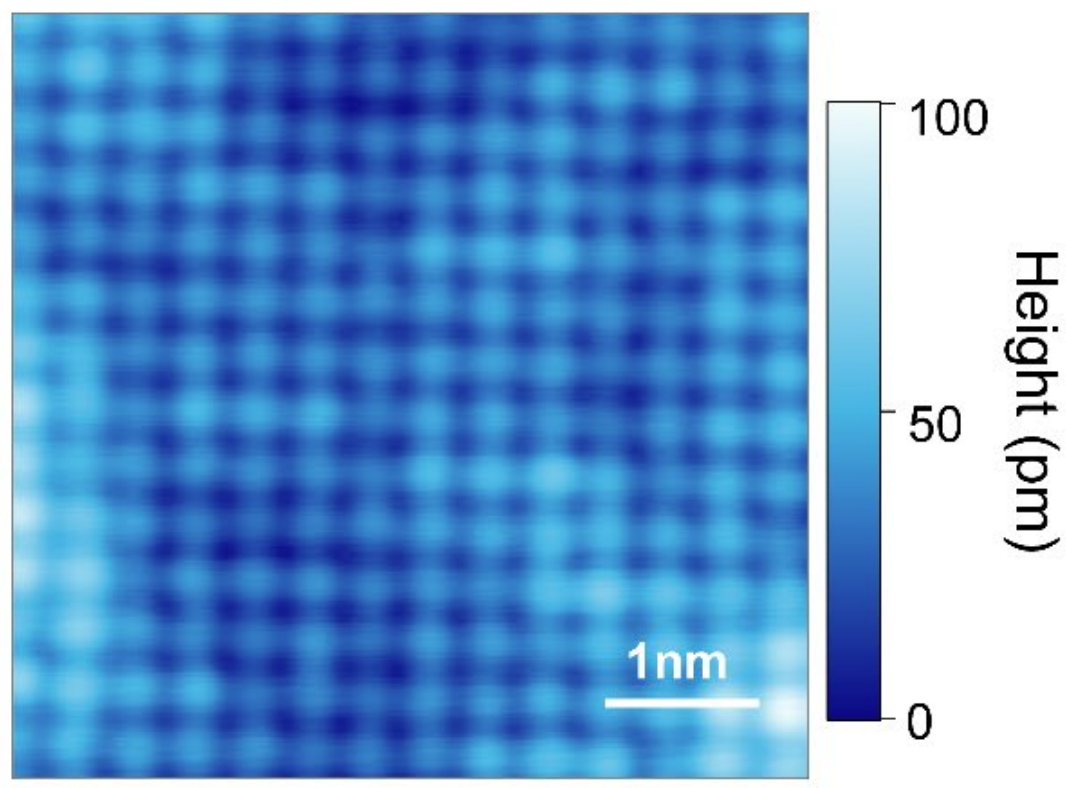

FIG. S1. Typical STM topography of FeSe films on STO (001) substrates (*S*15) showing the lattice structure. The sample bias $V_s$=80 mV and the tunneling current $I_t$=2.1 nA. The image is 5.6×5.6 $nm^2$.

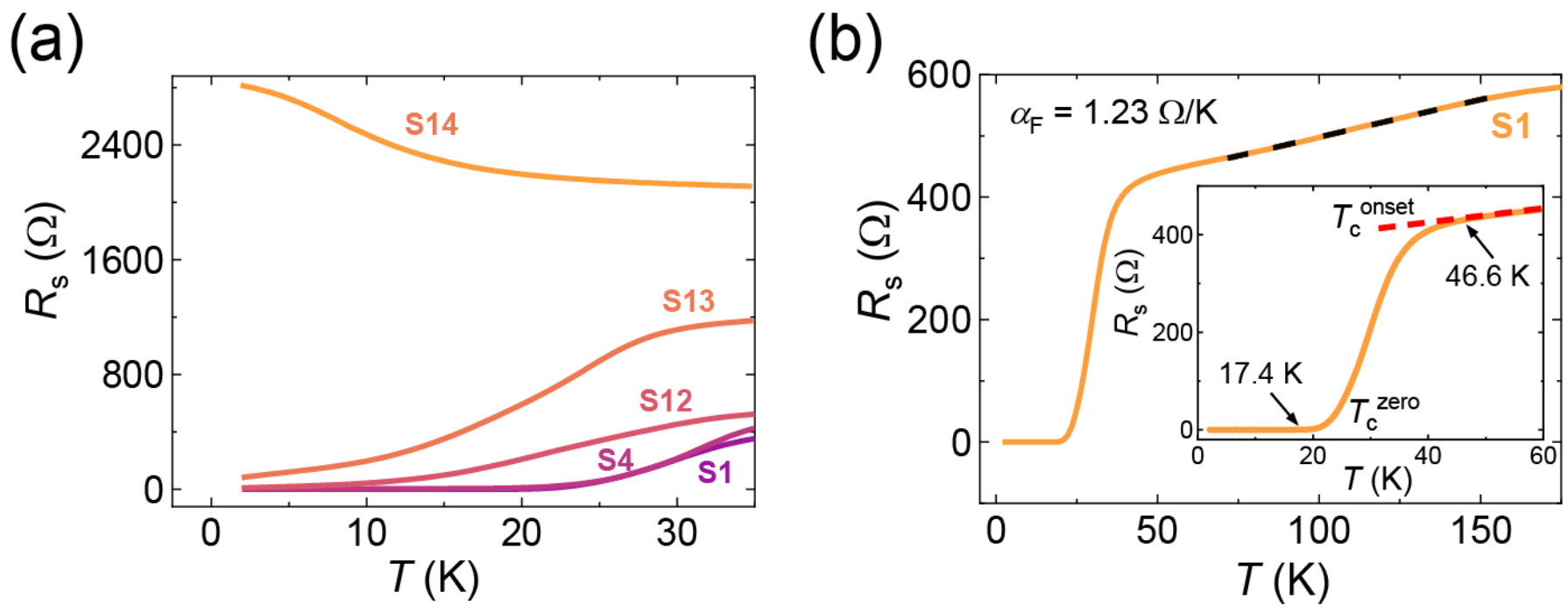

FIG. S2. Electric transport behavior of FeSe/STO at zero field. (a) Temperature dependent sheet resistance $R_s(T)$ of FeSe/STO samples ($S1$, $S4$, $S12$-$S14$), showing a superconductor to weakly localized metal transition with increasing normal state resistance. (b) The $R_s(T)$ curve of the superconducting FeSe/STO sample ($S1$) at zero magnetic field. The dashed black line is the linear fitting of the $R_s(T)$ curve above $T_c^{onset}$, yielding the slope $\alpha_F = 1.23\ \Omega/K$. Inset: The zoom in view of the same $R_s(T)$ curve for sample $S1$, showing $T_c^{onset}$ = 46.6 K and $T_c^{zero}$ = 17.4 K. The onset superconducting critical temperature is defined as the temperature where the resistance starts to deviate from the linear extrapolation of the normal state.

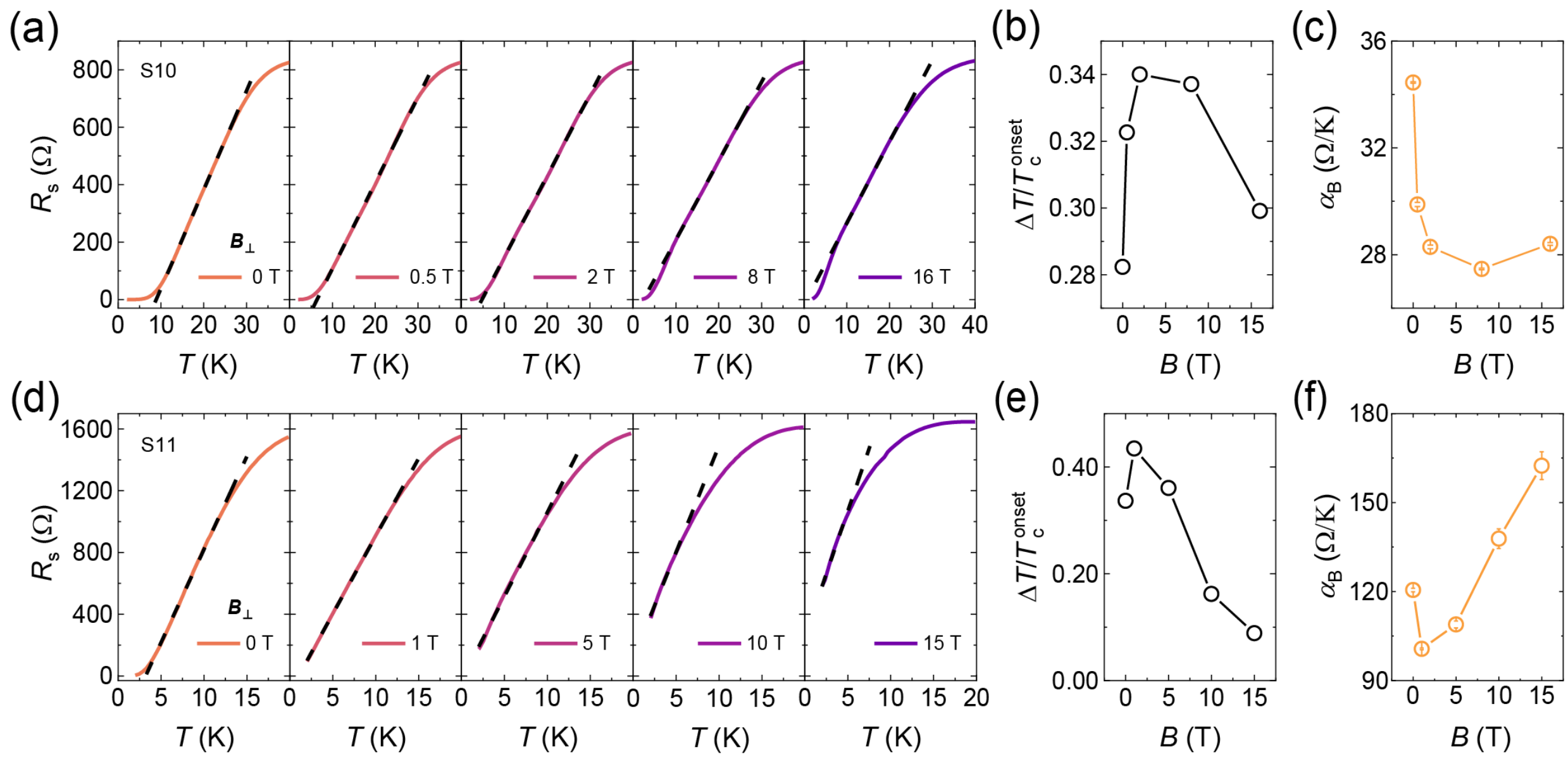

FIG. S3. $T$-linear resistance of FeSe/STO below $T_c^{\mathrm{onset}}$. (a, d) The $R_s(T)$ curves ((a) ($S10$) and (d) ($S11$)) under perpendicular magnetic fields. The black dashed lines are fits to the linear part of $R_s(T)$ curves, indicating a strange metal state. (b, e) Field dependence of the temperature ratio ($\Delta T/T_c^{\mathrm{onset}}$), $\Delta T$ is the temperature range of the linear $R_s(T)$ curves. (c, f) The slopes $\alpha_B$ of linear $R_s(T)$ under different fields in (a) and (d).

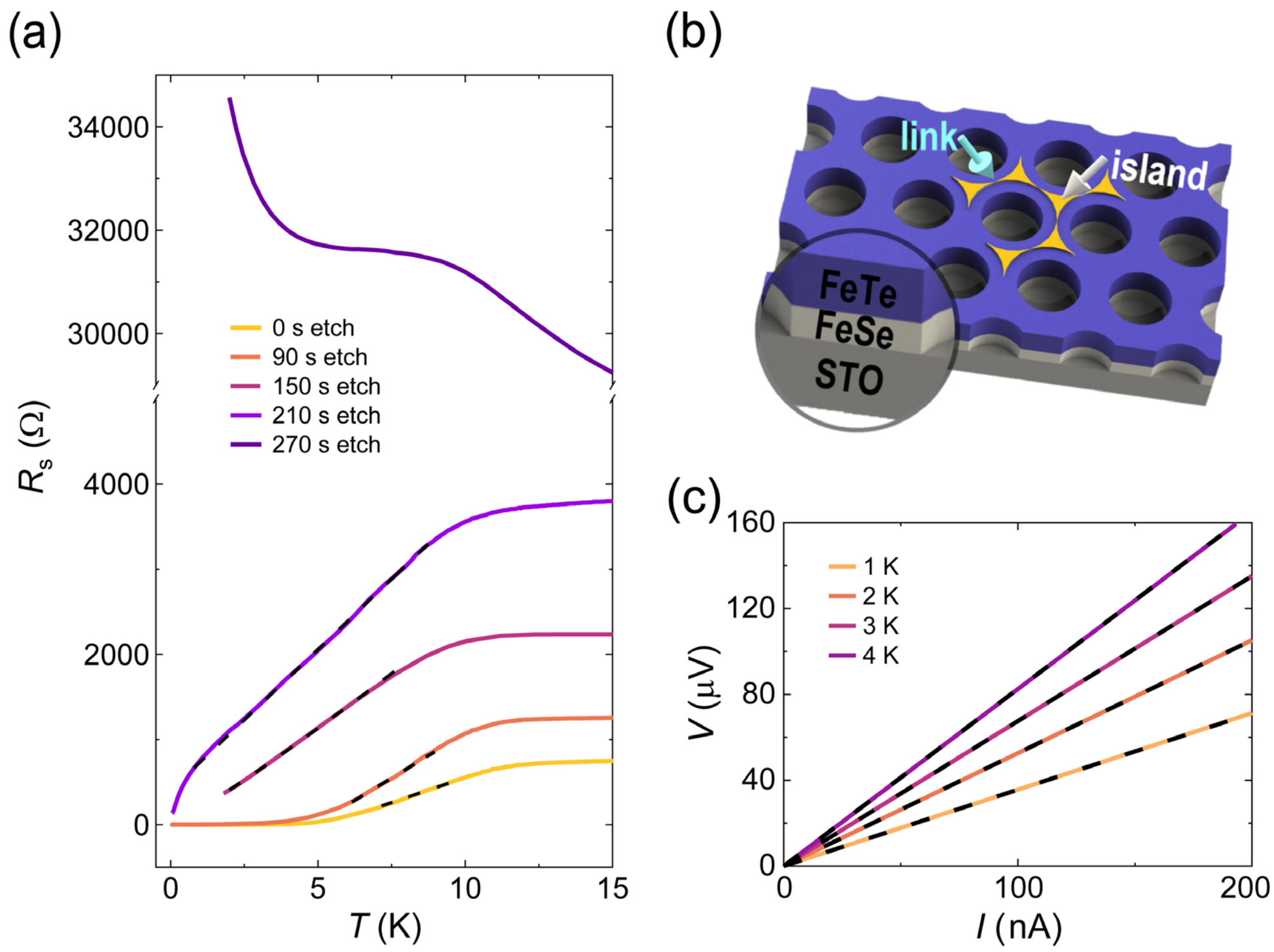

FIG. S4. $T$-linear resistance of nanopatterned FeSe/STO with triangular arrays of nano-holes (sample $S9$). (a) The $R_s(T)$ curves of FeSe films with different etching times. A shorter etching time corresponds to a smaller normal state resistance ($R_N$) and narrower range of linear $R_s(T)$. The linear $R_s(T)$ trend persists down to 1 K for 210 s etched film. The black dashed lines are linear fits to $R_s(T)$. (b) Schematic for nanopatterned FeSe film on STO substrate. Superconducting areas (marked as island) are connected by insulating links. (c), The $I$-$V$ curves showing Ohmic behavior within 200 nA down to 1 K for 210 s etched film.

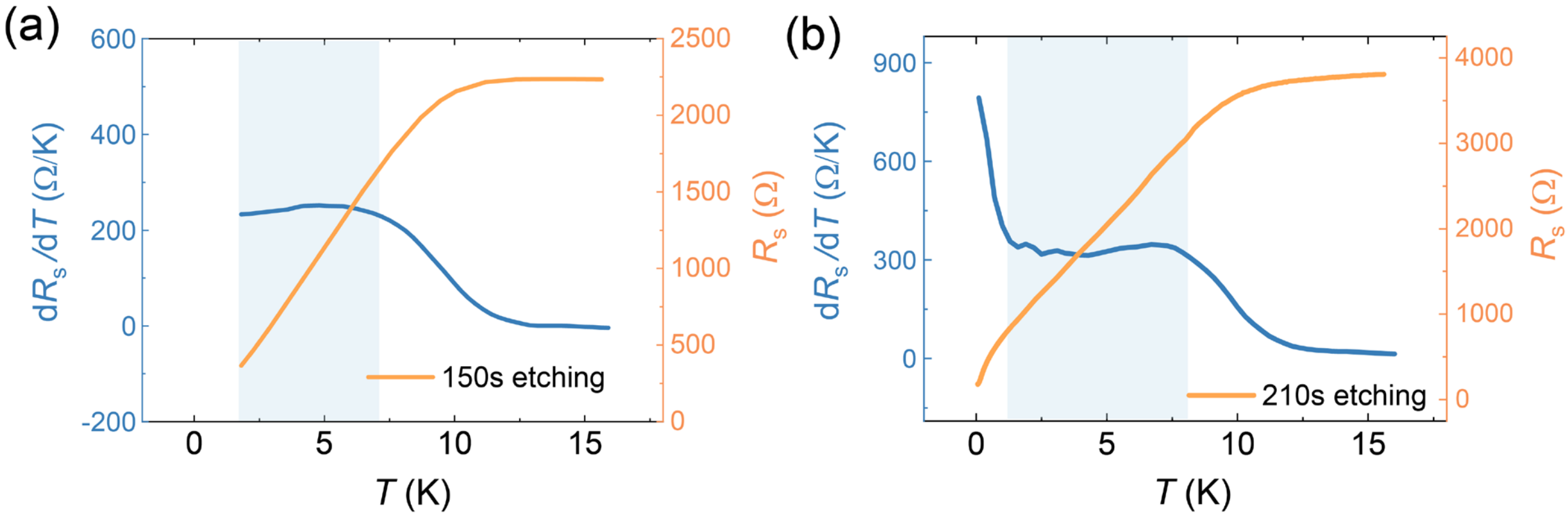

Fig. S5. $R_s$ versus $T$ curves and the corresponding differential resistance $dR_s/dT$ versus $T$ curves for the films ($S9$) with 150 s (a) and 210 s (b) etching time. The shadow areas are guides for the eye, where the plateaus are manifested in the $dR_s/dT$-$T$ curves corresponding to the linear $R_s(T)$ behaviors.

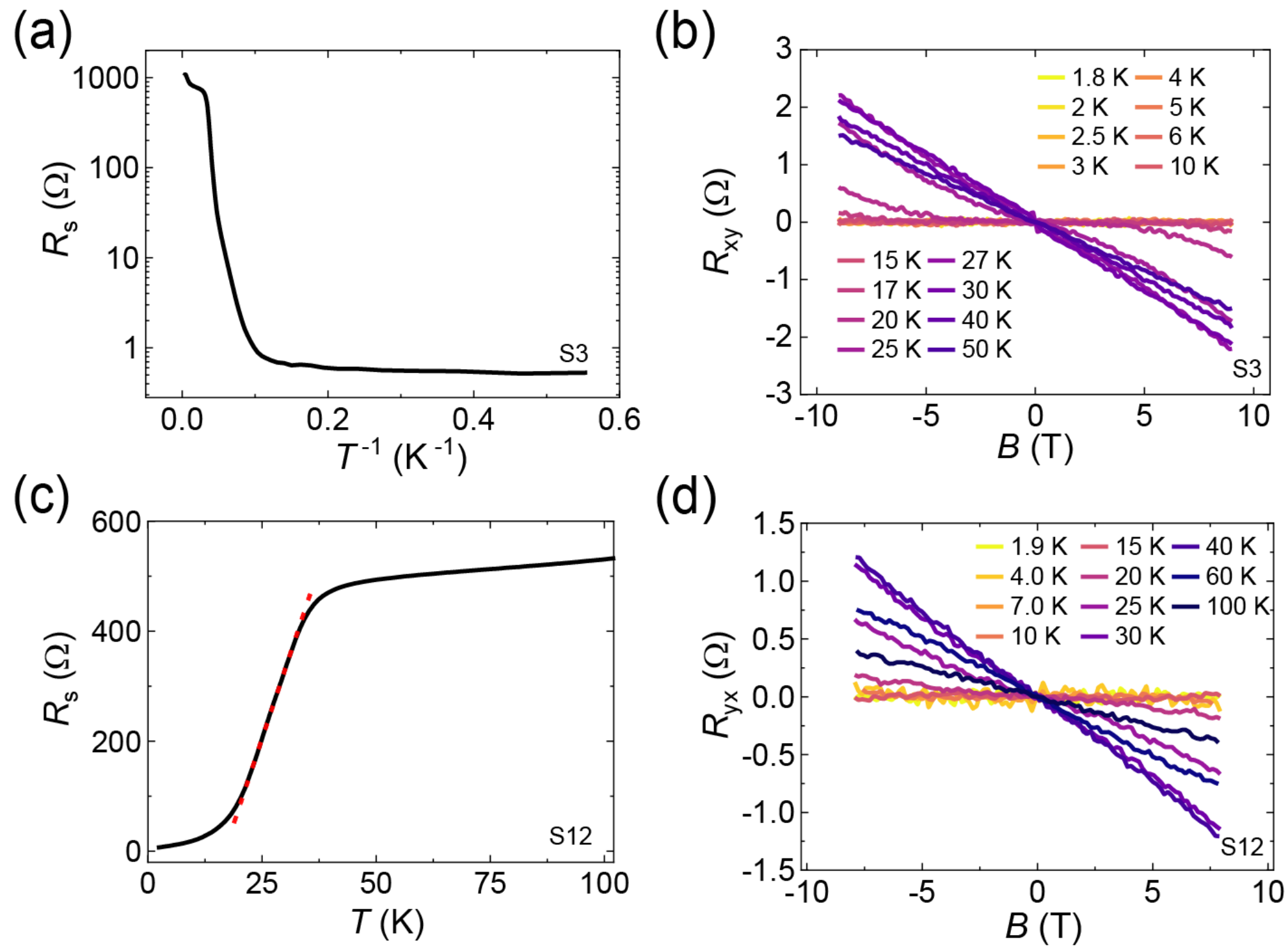

FIG. S6. $R_s(T)$ curves (a, c) and Hall resistance under perpendicular magnetic fields (b, d) of FeSe/STO. (a) and (b) are the experimental data for sample $S3$. (c) and (d) are the data for sample $S12$.

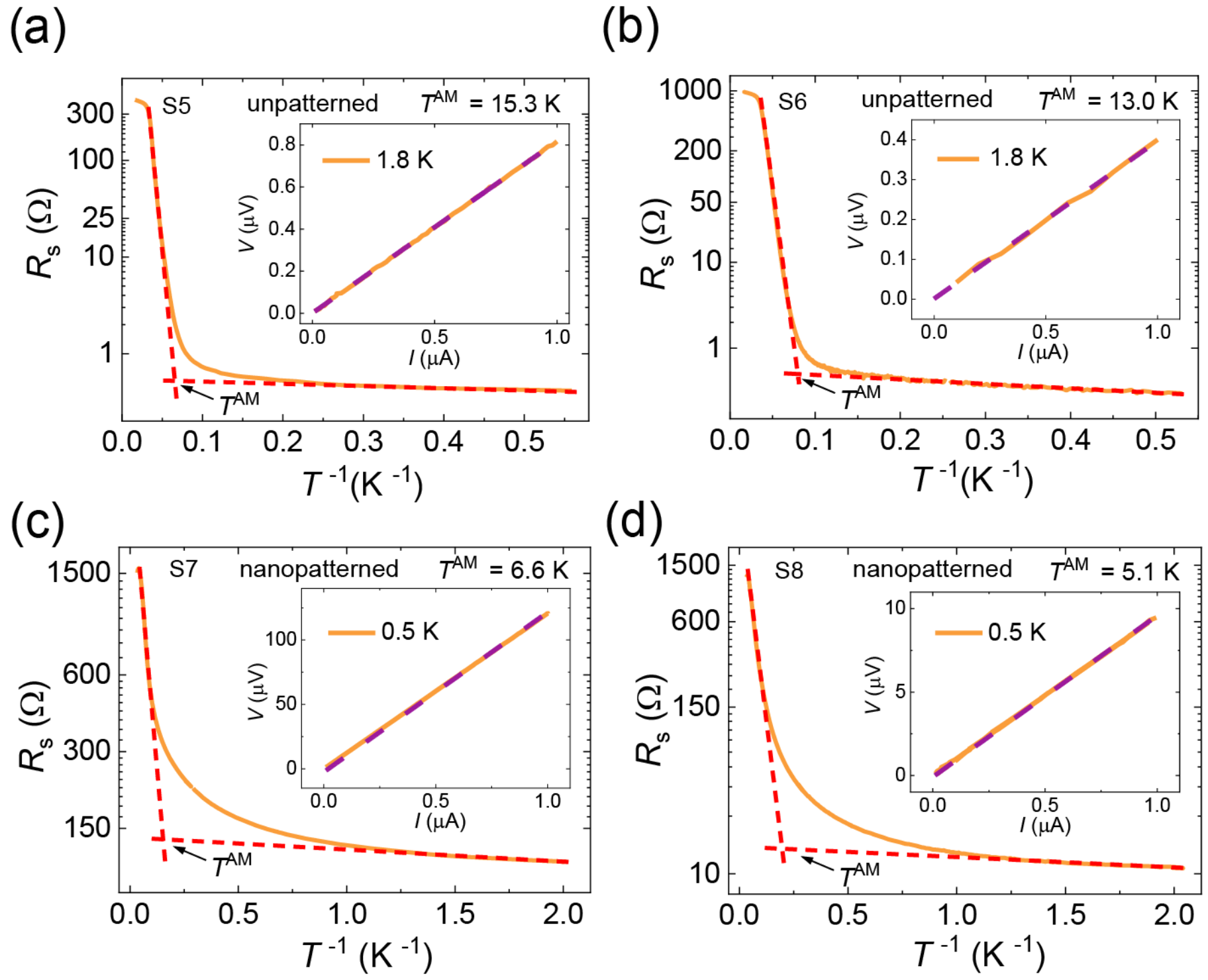

FIG. S7. Anomalous metal states in unpatterned (sample *S*5 (a), sample *S*6 (b)) and nanopatterned FeSe/STO (sample *S*7 (c), sample *S*8 (d)). Insets are the linear *I*-*V* curves at 1.8 K for (a), (b) and 0.5 K for (c), (d), respectively.

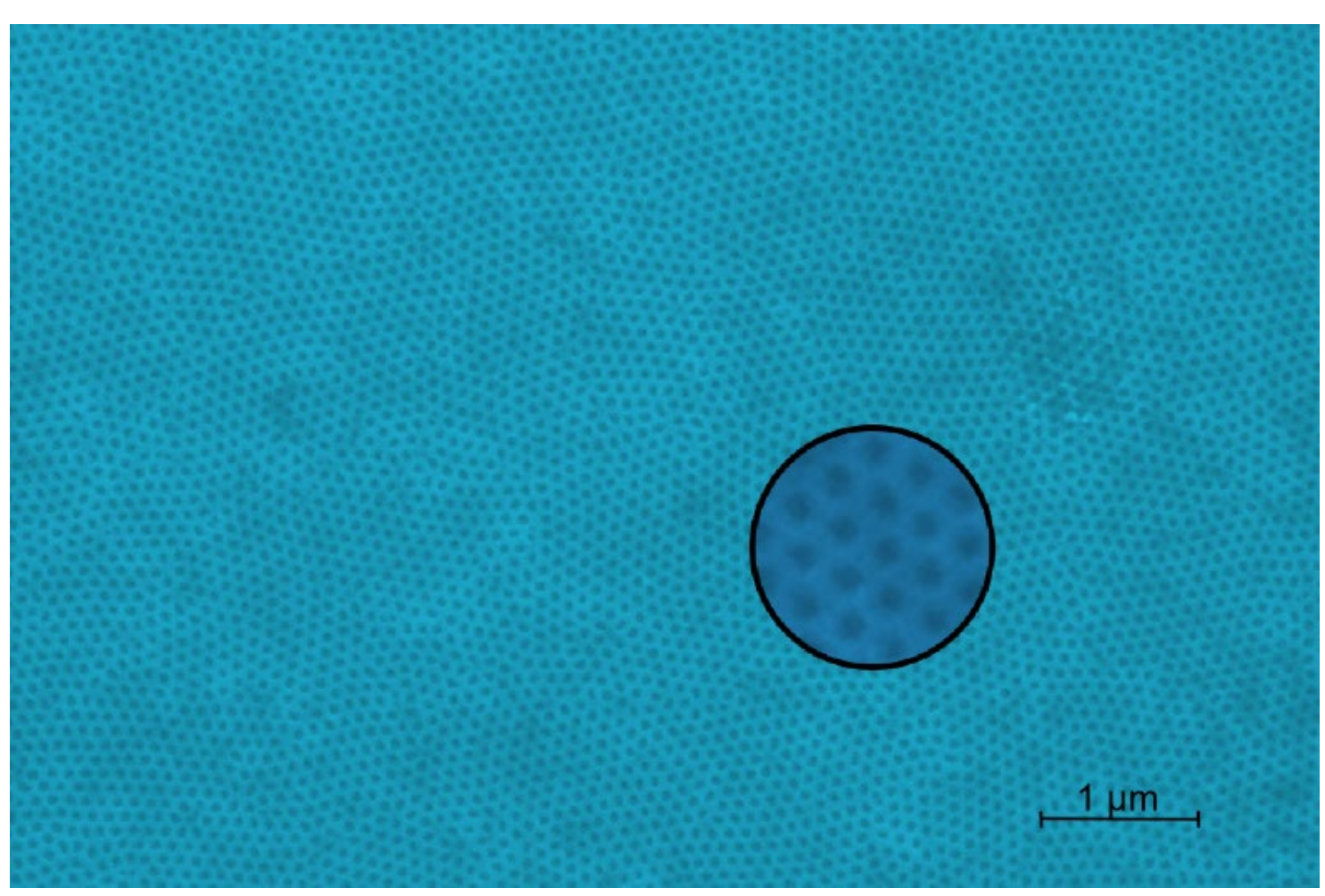

FIG. S8. Typical scanning electron microscope image of a nanopatterned FeSe film. The inset is the zoom-in of the nanopatterned structure.

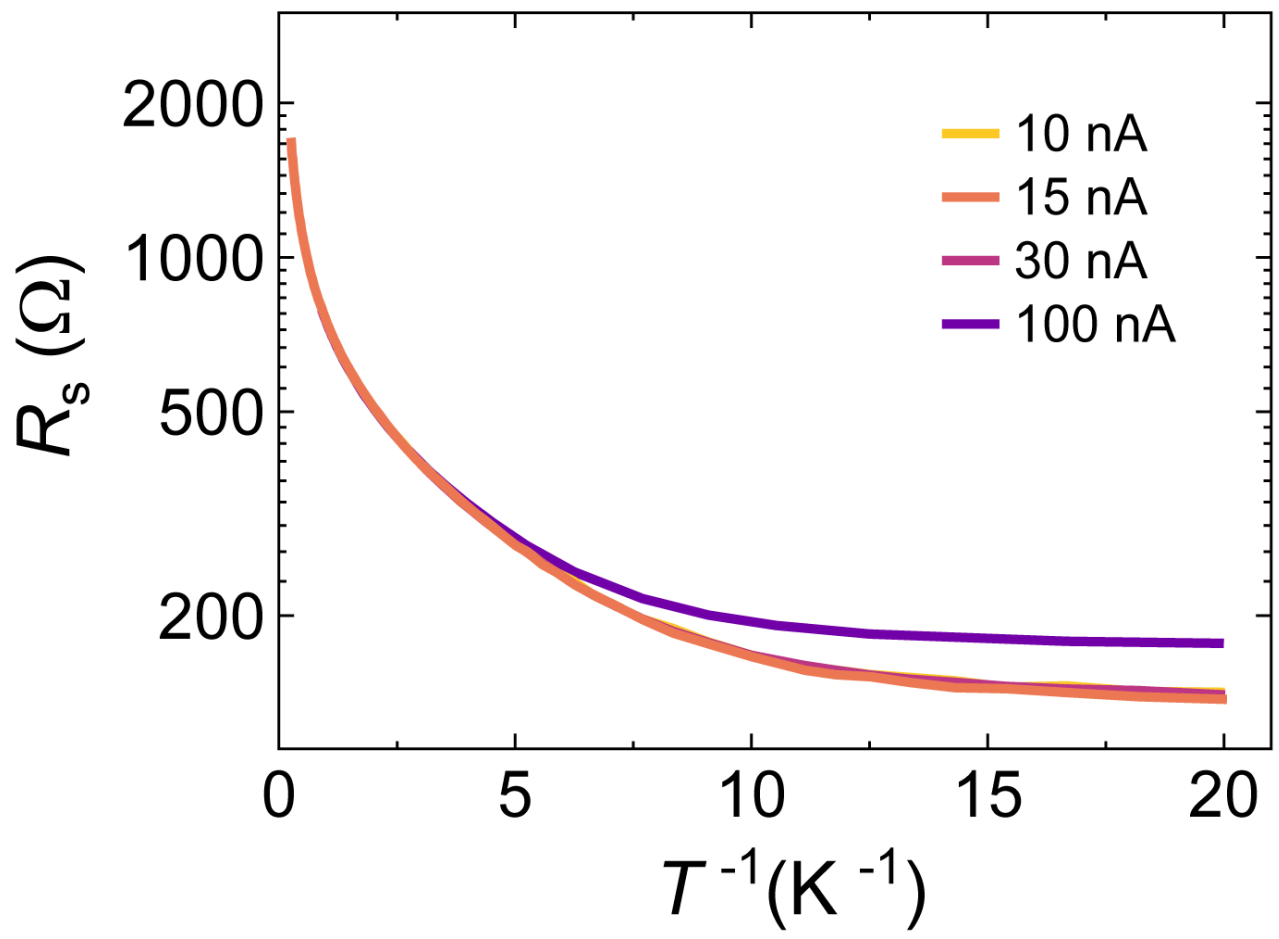

FIG. S9. Arrhenius plots of $R_s(T)$ curves under different currents for FeSe film ($S$9-210 s etch). The low-temperature saturated resistance in $R_s(T)$ curves becomes larger when measured with 100 nA, a current beyond the Ohmic regime.

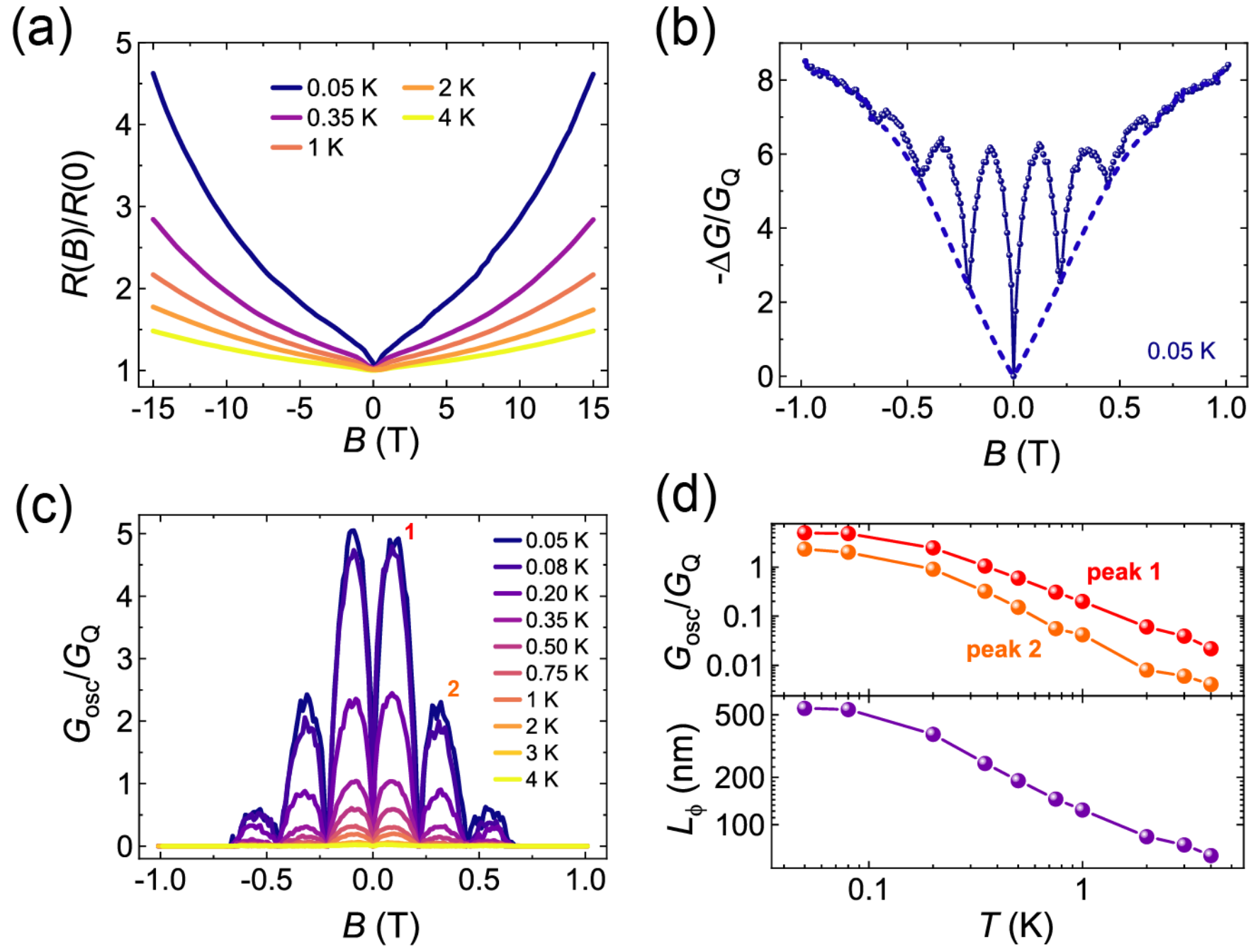

FIG. S10. Magneto-resistance and background subtraction of magnetoconductance oscillations of 210 s etched film ($S9$) showing anomalous metal state. (a) Giant positive magnetoresistance at various temperatures from -15 T to 15 T. (b) Negative change of magnetoconductance $-\Delta G/G_{\mathrm{Q}}$ at 0.05 K. $G_{\mathrm{Q}}$ is the quantum conductance of Cooper pairs ($\frac{4e^2}{h}$). $-\Delta G = G_0 - G(B)$, where $G_0$ is the conductance at zero magnetic field. The dashed line is the polynomial fitting of the magnetoconductance background. (c) Magnetoconductance oscillations at various temperatures. $G_{\mathrm{osc}}$ is the amplitude of conductance after subtracting the background. (d) The amplitude of magnetoconductance oscillations $G_{\mathrm{osc}}$ (upper panel, the corresponding peak levels are indicated in (c)) and the phase coherence length $L_{\phi}$ (lower panel, derived from magnetoconductance oscillations of peak 2) as a function of temperature.

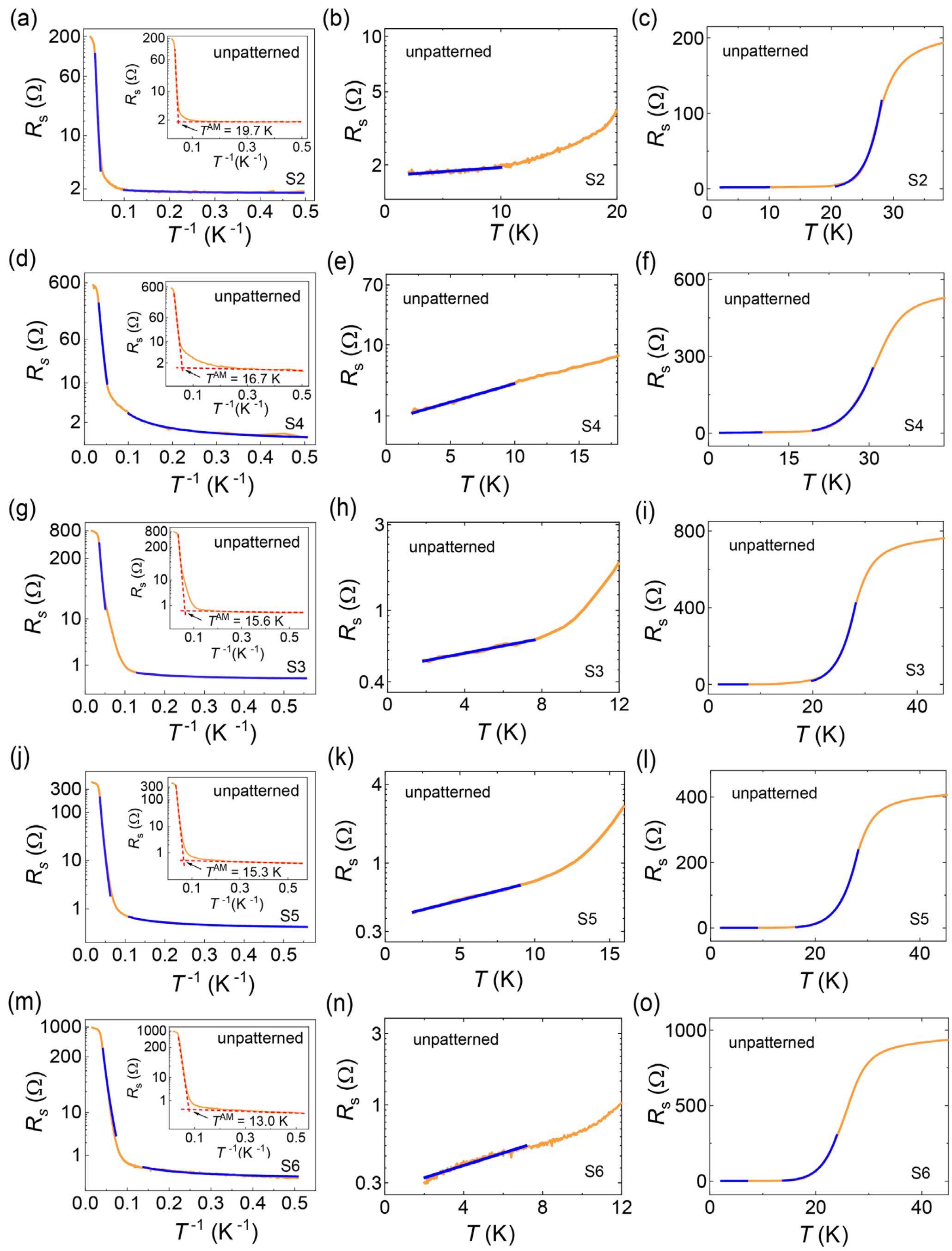

FIG. S11. The $R_s(T)$ curves and theoretical fitting results with Eq. (SE15) in unpatterned FeSe/STO samples showing anomalous metal state. (a, d, g, j, m) The $R_s(T)$ curves in lg$R_s$ vs $T^{-1}$ plot. The insets indicate the definition of $T^{AM}$. (b, e, h, k, n) The $R_s(T)$ curves in lg$R_s$ vs $T$ plot, in which the linear fittings verify that $R_s \sim \exp(T/T_0)$ at low temperatures. (c, f, i, l, o) The $R_s(T)$ curves in $R_s$ vs $T$ plot. The power law temperature dependence of resistance ($R_s \propto T^{2\bar{\gamma}-1}$, $\bar{\gamma}$ is the average dissipation strength) is verified in the moderate temperature regime. In all the panels, the orange lines are the experimental $R_s(T)$ curves and the blue lines represent the fitting results with our theoretical model. The fitting parameters are shown in Table S1.

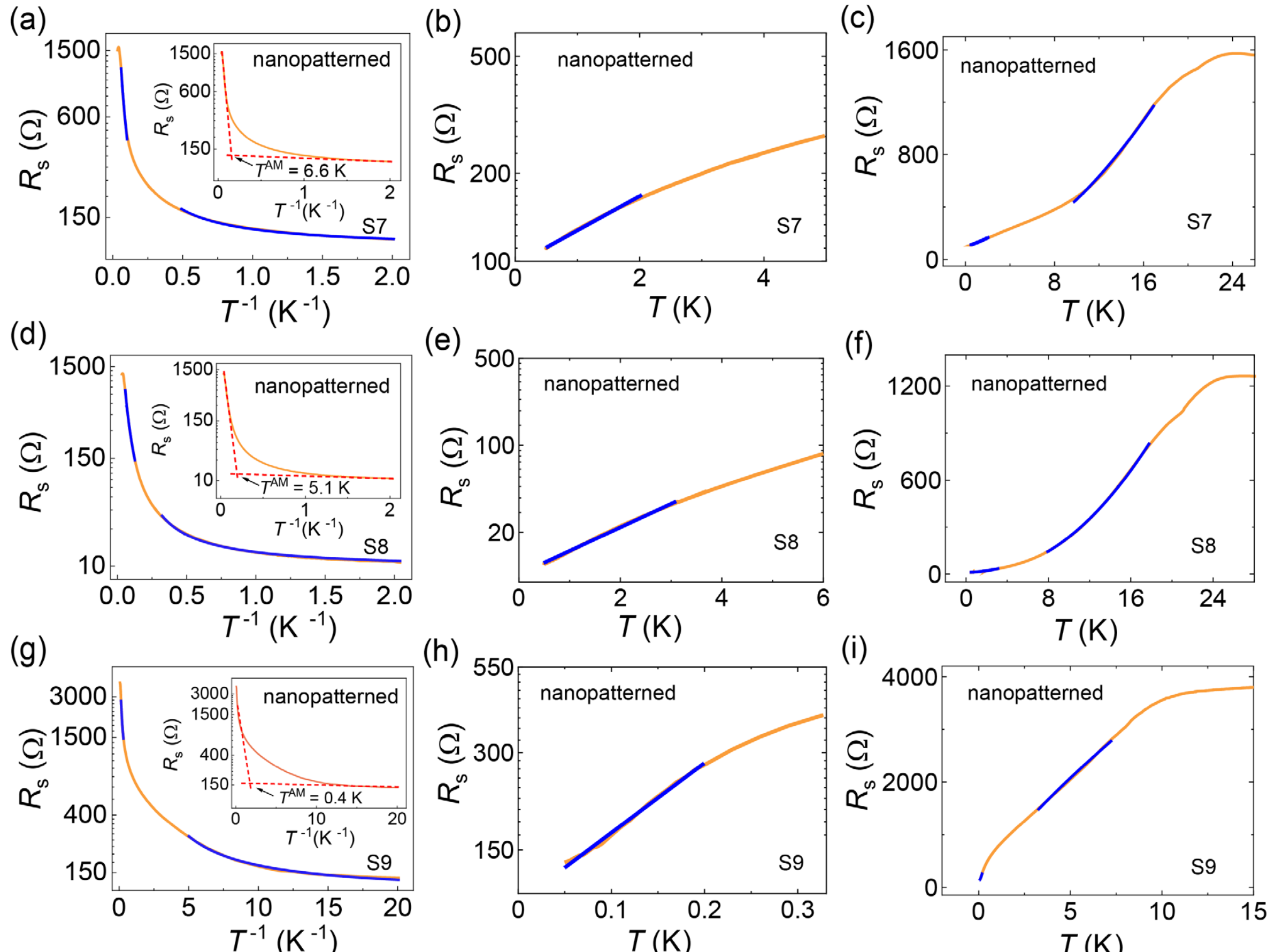

FIG. S12. The $R_s(T)$ curves and fitting results with Eq. (SE15) of anomalous metal state in nanopatterned FeSe/STO samples. (a, d, g) The $R_s(T)$ curves in lg$R_s$ vs $T^{-1}$ plot. The insets indicate the definition of $T^{AM}$. (b, e, h) The $R_s(T)$ curves in lg$R_s$ vs $T$ plot, in which the linear fittings verify that $R_s \sim \exp(T/T_0)$ at low temperatures. (c, f, i) The $R_s(T)$ curves in $R_s$ vs $T$ plot. The power law temperature dependence of resistance ($R_s \propto T^{2\bar{\gamma}-1}$, $\bar{\gamma}$ is the average dissipation strength) is verified in the moderate temperature regime. In all the panels, the orange lines are the experimental $R_s(T)$ curves and the blue lines represent the fitting results with our theoretical model. The fitting parameters are shown in Table S1.

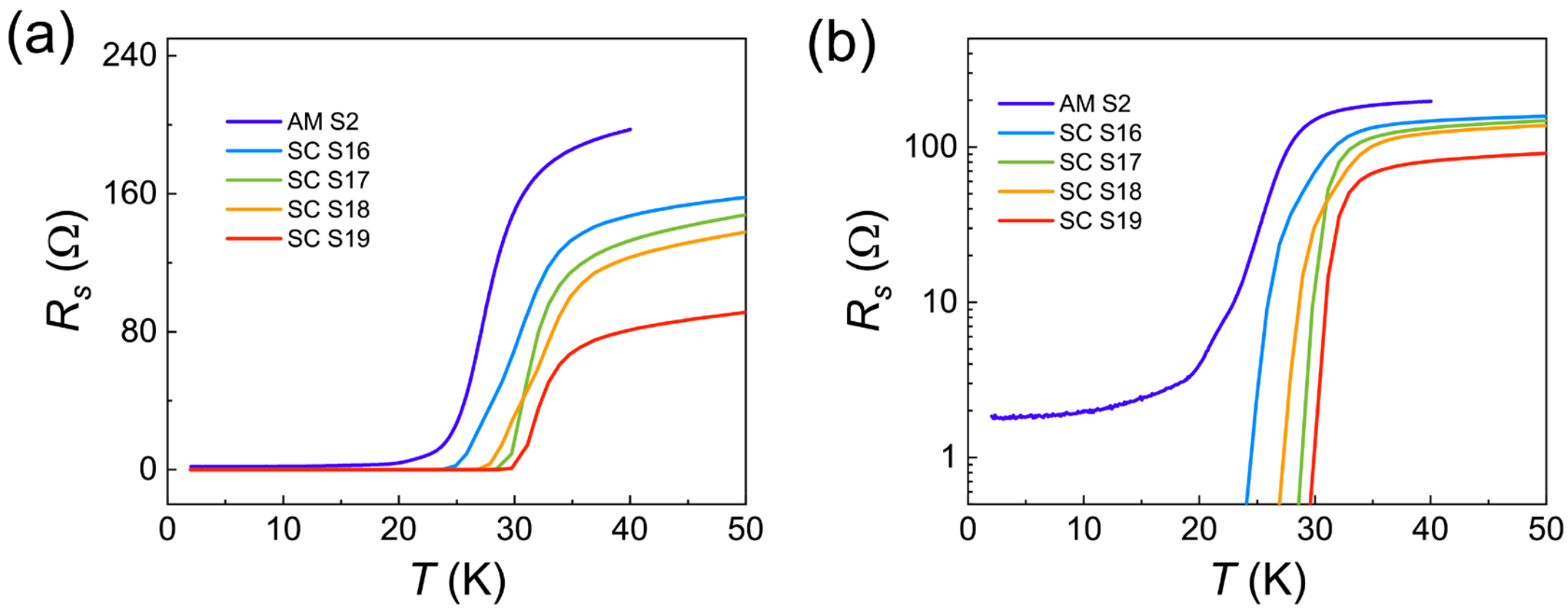

Fig. S13. $R_s$-$T$ curves of different FeSe/STO samples with different $R_N$ in the linear-$R_s$-scale (a) and log$R_s$-scale (b). Here, SC or AM denotes that the sample shows superconducting state or anomalous metal state at low temperatures.

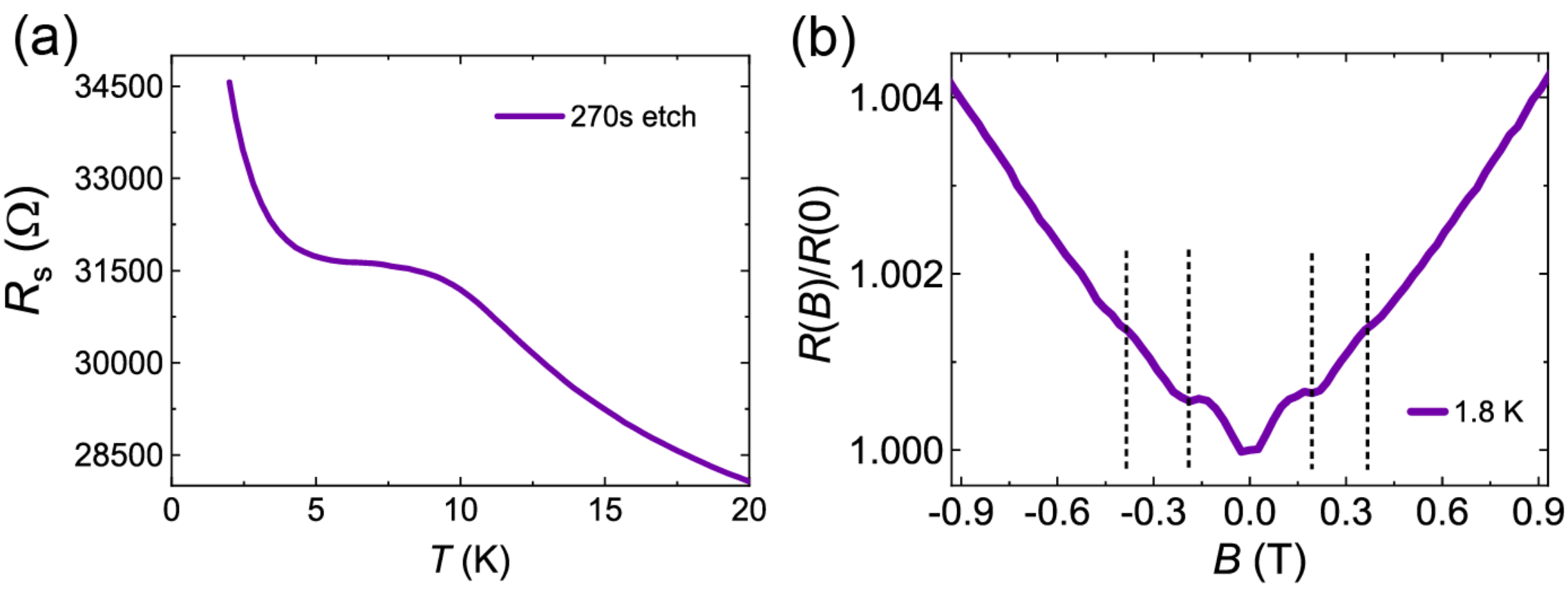

FIG. S14. Insulating state of nanopatterned FeSe/STO ($S$9-270 s etch). (a) $R_{\mathrm{s}}(T)$ curve of 270 s etched FeSe film, showing a bosonic insulating state. In the framework of superconductor-insulator transitions, normally the bosonic insulator refers to the samples with resistance larger than the sheet resistance of quantum critical point, which is the quantum resistance of Cooper-pairs $(h/(2e)^2$ ≈6.45 kΩ)[58]. The sheet resistance of 270 s etched film is around 34.6 kΩ at 1.8 K, which is much larger than 6.45 kΩ. (b) Magnetoresistance of 270 s etched FeSe film at 1.8 K. Magnetoresistance oscillations with period ~0.215 T are observed, corresponding to one superconducting flux quantum per unit cell of the nanopattern, which indicates the bosonic nature of the insulating state.

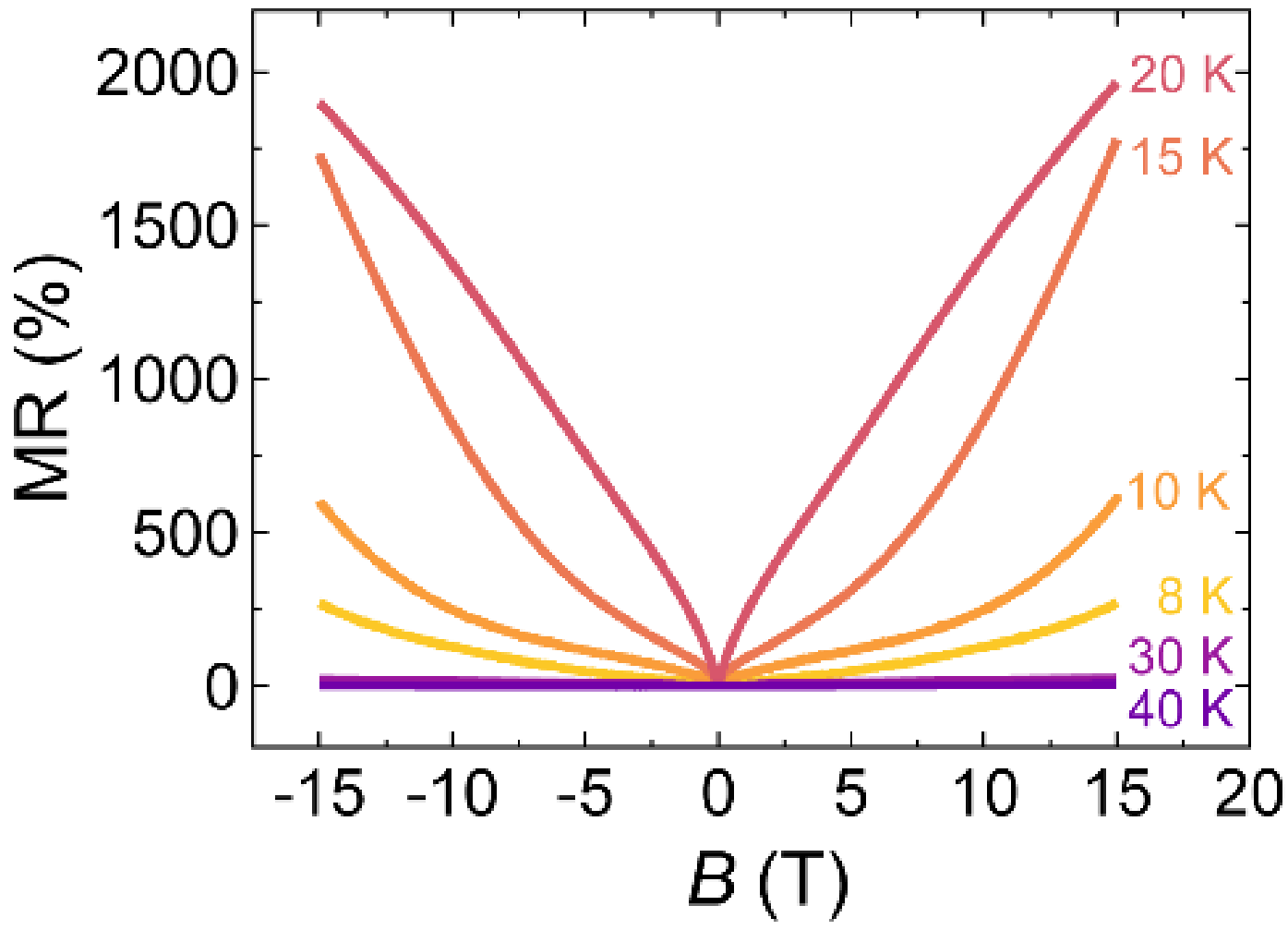

FIG. S15. Magnetoresistance of the high-temperature anomalous metal state and normal state in FeSe/STO (*S*2). The magnetoresistance (MR=($R(B)$-$R(0)$)/$R(0)$×100%) is 270%-1900% below $T^{\mathrm{AM}}$ (~20 K) at 15 T. Above $T_{\mathrm{c}}^{\mathrm{onset}}$, the magnetoresistance is only 3.5% at 40 K and 15 T.

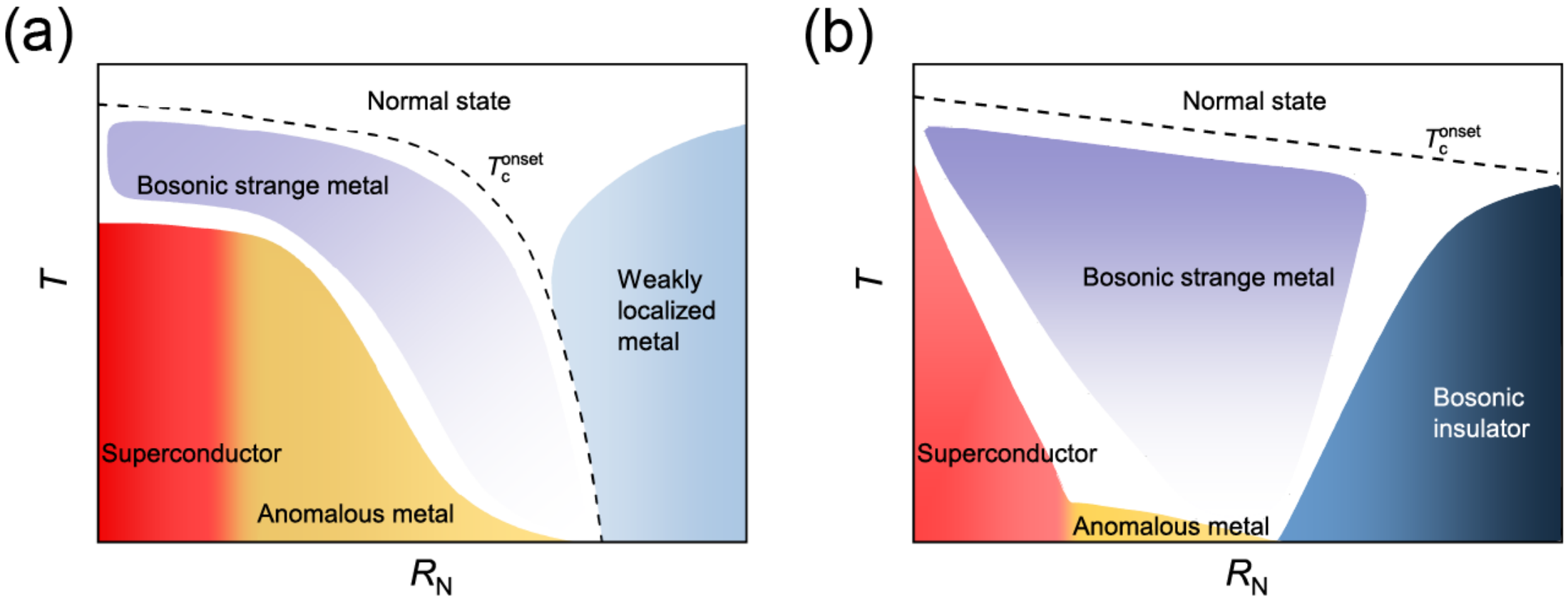

FIG. S16. The schematic phase diagrams of unpatterned FeSe/STO (a) and nanopatterned FeSe/STO samples (b). Below $T_c^{onset}$, superconductor, anomalous metal, and bosonic strange metal are characterized by zero resistance within the instrument resolution, residual resistance plateau, and $T$-linear resistance, respectively. Weakly localized metal and insulator show increasing resistance as $T \rightarrow 0$. With increasing $R_N$, the unpatterned FeSe/STO samples in (a) show superconductor-anomalous metal transition. The anomalous metal can persist up to 19.7 K, comparable to the zero-resistance superconducting transition temperature. With further increasing $R_N$, the unpatterned FeSe/STO becomes a weakly localized metal. The nanopatterned FeSe/STO samples in (b) show superconductor-anomalous metal-insulator transitions with increasing $R_N$. The temperature regime of anomalous metal state is relatively small, while the bosonic strange metal state occupies a large area in the phase diagram. The $\frac{h}{2e}$ quantum oscillations exist in the anomalous metal (Fig. 2(d)), bosonic strange metal (Fig. 2(d)), and insulating (Fig. S14) states in nanopatterned FeSe/STO, indicating that Cooper pairs dominate the transport of these states.

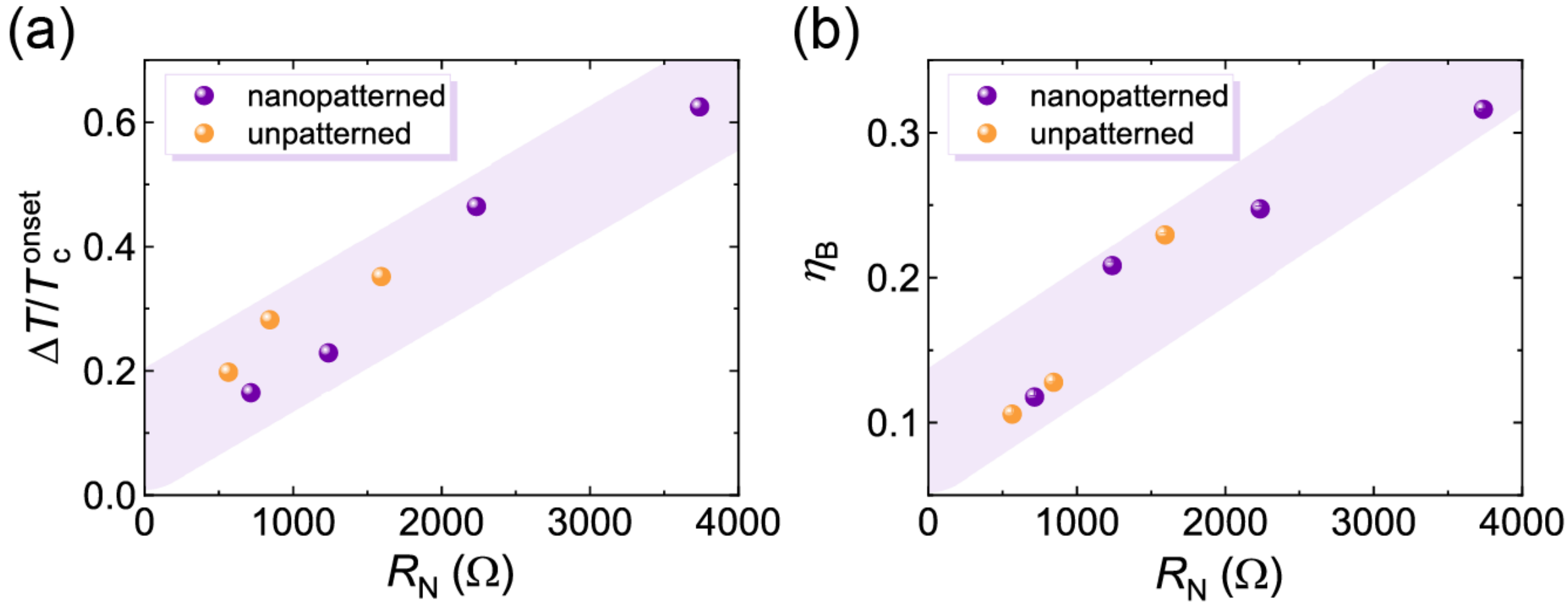

Fig. S17. $\Delta T/T_c^{onset}$ ($\Delta T$ is the temperature regime of the $T$-linear resistance) (a), and $\eta_B$ (b) as a function of normal state resistance ($R_N$) at zero field for unpatterned and nanopatterned FeSe/STO. Both $\Delta T/T_c^{onset}$ and $\eta_B$ exhibit positive correlation with $R_N$. The shadow areas are guides for the eye. According to the previous works, a dimensionless coefficient $\eta_F = \frac{2e^2}{h} \cdot \alpha_F \cdot T_F$ was defined for the slope $\alpha_F$ of fermionic strange metal, where $T_F$ is the Fermi temperature and $\eta_F$ is around 0.2-1.6[4,10]. Here, in our system, the $T$-linear resistance manifests below $T_c^{onset}$, allowing us to define an effective coefficient $\eta_B = \frac{2e^2}{h} \cdot \alpha_B \cdot T_c^{onset}$ by analogy, where $T_c^{onset}$ represents the characteristic energy scale for Cooper pairs. The typical value of $\eta_B$ is around 0.1-0.3 for our FeSe/STO samples showing $T$-linear resistance.

**Table. S1. Fitting parameters of $R_s(T)$ curves with our theoretical model for anomalous metal states.** [$T_1$ (K), $T_2$ (K)] represents the temperature range for fitting. The fitting curves are shown in Fig. S11 and S12. Here, $T^*$ is chosen to be near $T_c^{\text{onset}}$. These tables only involve the samples showing anomalous metallic states, instead of all measured samples.

| | | $R_s = A \cdot \left(\frac{T}{T^*}\right)^{2\bar{\gamma}-1}$, $T^* = 35$ K | | |
|---|---|---|---|---|
| Sample | Thickness (UC) | $A$ (kΩ) | $2\bar{\gamma}-1$ | [$T_1$(K), $T_2$ (K)] |
| $S2$ | 5 | $1.61 \pm 0.06$ | $11.9 \pm 0.1$ | [20.6, 28.1] |
| $S3$ | 1 | $3.0 \pm 0.1$ | $9.1 \pm 0.1$ | [19.7, 28.2] |
| $S4$ | 1 | $0.64 \pm 0.01$ | $7.10 \pm 0.08$ | [19.3, 30.7] |
| $S5$ | 1 | $1.50 \pm 0.03$ | $8.54 \pm 0.07$ | [16.2, 28.2] |
| $S6$ | 1 | $7.0 \pm 0.1$ | $8.27 \pm 0.03$ | [13.7, 24.0] |
| $S7$ | 2 | $4.3 \pm 0.2$ | $1.79 \pm 0.05$ | [9.8, 16.9] |
| $S8$ | 2 | $3.65 \pm 0.04$ | $2.18 \pm 0.01$ | [7.9, 17.8] |
| $S9$-210 s etch | 2 | $9.7 \pm 0.1$ | $0.800 \pm 0.007$ | [3.2, 7.3] |

| | | $R_s = R_0 \cdot Exp(\frac{T}{T_0})$ | | |
|---|---|---|---|---|
| Sample | Thickness (UC) | $R_0$ (Ω) | $T_0$ (K) | [$T_1$(K), $T_2$ (K)] |
| $S2$ | 5 | $1.744 \pm 0.008$ | $95 \pm 6$ | [2.0, 10.1] |
| $S3$ | 1 | $0.48 \pm 0.06$ | $21 \pm 1$ | [1.8, 7.7] |
| $S4$ | 1 | $0.86 \pm 0.01$ | $8.3 \pm 0.1$ | [2.0, 10.0] |
| $S5$ | 1 | $0.37 \pm 0.02$ | $15.2 \pm 0.2$ | [1.8, 9.0] |
| $S6$ | 1 | $0.27 \pm 0.02$ | $10.4 \pm 0.1$ | [2.0, 7.2] |
| $S7$ | 2 | $97.7 \pm 0.3$ | $3.74 \pm 0.03$ | [0.5, 2.0] |
| $S8$ | 2 | $9.19 \pm 0.03$ | $2.297 \pm 0.008$ | [0.5, 3.1] |
| $S9$-210 s etch | 2 | $103 \pm 1$ | $0.202 \pm 0.002$ | [0.05, 0.2] |

**Table S2. The power law fitting ($R_s(T) = a + bT^n$) of $R_s(T)$ curves for bosonic strange metal states at various magnetic fields or with different etching time.** [$T_1$ (K), $T_2$ (K)] denotes the temperature regime of the $T$-linear resistance.

| ***S*9** | **etching time** | **210 s** | **150 s** | **90 s** | **0 s** | |
|---|---|---|---|---|---|---|
| | n | $0.989 \pm 0.053$ | $0.993 \pm 0.017$ | $1.141 \pm 0.032$ | $0.992 \pm 0.012$ | |
| | [$T_1$ (K), $T_2$ (K)] | [1.0, 8.7] | [1.8, 8.7] | [6.2, 9.0] | [7.2, 9.2] | |
| ***S*10** | $B_\perp$ | **0 T** | **0.5 T** | **2 T** | **8 T** | **16 T** |
| | n | $0.976 \pm 0.002$ | $1.046 \pm 0.002$ | $1.059 \pm 0.002$ | $1.006 \pm 0.003$ | $0.976 \pm 0.003$ |
| | [$T_1$ (K), $T_2$ (K)] | [3.3, 26.9] | [11.7, 26.5] | [11.1, 26.3] | [9.2, 24.3] | [8.5, 21.9] |
| ***S*11** | $B_\perp$ | **0 T** | **1 T** | **5 T** | **10 T** | **15 T** |
| | n | $0.972 \pm 0.014$ | $0.964 \pm 0.004$ | $0.884 \pm 0.087$ | $0.853 \pm 0.010$ | $0.989 \pm 0.079$ |
| | [$T_1$ (K), $T_2$ (K)] | [3.6, 11.9] | [2.0, 11.6] | [2.0, 10.5] | [2.1, 5.6] | [2.0, 3.9] |